\documentclass[sigconf]{acmart}

\usepackage{diagbox}
\usepackage{subfig}
\RequirePackage{algorithm, algorithmic}

\AtBeginDocument{%
  }

\setcopyright{none}
\copyrightyear{}
\acmYear{}
\acmDOI{}
\acmConference{}
\acmBooktitle{}
\acmISBN{}

\begin{document}

\title{A Heuristic Algorithm for Shortest Path Search}

\author{Huashan Yu}
\affiliation{%
  \institution{Peking University}
  \city{Beijing}
  \country{China}
}
\email{yuhs@pku.edu.cn}

\author{Xiaolin Wang}
\affiliation{%
  \institution{Peking University}
  \city{Beijing}
  \country{China}
}
\email{wxl@pku.edu.cn}

\author{Yingwei Luo}

\affiliation{%
  \institution{Peking University}
  \city{Beijing}
  \country{China}
}
\email{lyw@pku.edu.cn}


\begin{abstract}
The Single-Source Shortest Path (SSSP) problem is well-known for the challenges in developing fast, practical, and work-efficient parallel algorithms. This work introduces a novel shortest path search method. It allows paths with different lengths to be extended in parallel at the cost of almost negligible repeated relaxations. A dynamic-stepping heuristic is proposed for the method to efficiently reduce the extended paths and the synchronizations. A traversal-optimization heuristic is proposed to improve the method by efficiently reducing the created paths and alleviating the load imbalance. Based on the method, the two heuristics are used to develop a practical SSSP algorithm, which tactfully reduces workload and overhead. The heuristics and the algorithm were evaluated on 73 real-world and synthetic graphs. The algorithm was also compared with five state-of-the-art SSSP implementations. On each GAP benchmark suite graph except Road, its speedup to the best achieved by these five implementations is 2.5x to 5.83x.
\end{abstract}

%

%

\maketitle

\section{Introduction}
Graphs are a core part of most analytics workloads. The graph abstraction uses edges to represent the relationships among real-world entities, and enables graph algorithms to analyze networks with different structures and from different domains in the same way. The Single-Source Shortest Path (SSSP) problem is a fundamental graph primitive. It computes the shortest paths from a source vertex to all reachable vertices in a weighted graph. The problem has numerous practical and theoretical applications, including combinatorial optimization~\cite{1, 2}, complex network analysis~\cite{3, 4, 5}, etc.  

SSSP is well-known for the challenges in developing fast, practical and work-efficient parallel algorithms. It has been selected as a core kernel in both Graph500 benchmark~\cite{6} and GAP benchmark suite (GAPBS)~\cite{7}. Typically, an SSSP computation is solved by relaxing the input graph's edges concurrently to minimize every vertex’s tentative distance. Here, a vertex's tentative distance represents the length of the shortest known path from the source vertex, initialized to $+\infty$ for unvisited vertices. Every edge relaxation creates a new path by extending an existing path with the relaxed edge. Let $v$ denote the vertex reached by the new path and $npl$ denote the new path's length. The relaxation updates $v$'s tentative distance to $npl$ if $v$'s current tentative distance is greater than $npl$. The relaxation is \emph{redundant} if the new path is longer than the shortest path between $v$ and the source vertex. The computation is completed after every vertex's tentative distance has been finalized. Its progress is checked iteratively by synchronizing the edge relaxations executed in the same round. 
 
In most large real-world graphs, the vertices are connected randomly, and the edges are several to tens of times of the vertices~\cite{8, 9, 10,11,12,13,14}. In turn, there are usually numerous cycles in a graph, and vertices with higher degrees tend to participate in more cycles~\cite{15,16, 17, 18}. Each cycle has one or more edges that are \emph{skippable} during the SSSP computation. An edge is skippable if every relaxation on it is redundant. Relaxations on the skippable edges decrease the SSSP computation efficiency. They introduce extra communication overhead and exacerbate workload imbalance in parallel settings. At the same time, the parallel paths are usually varied in the lengths. Due to the skippable edges, it is difficult for the parallel paths with different lengths to share the synchronization overhead without the sacrifice in efficiency. This kind of overhead sharing often causes many edges to be repeatedly relaxed.

Distribution of the skippable edges is statistically correlated with the input graph’s vertex degrees and edge weights. This work proposes to optimize SSSP computations on undirected graphs with vertex-degree statistics and edge-weight statistics. Our contributions mainly include:
\begin{itemize}
\item{A novel shortest path search method that allows the paths with different lengths and different edge numbers to be extended in parallel at the cost of almost negligible repeated edge relaxations.} It guarantees a subtle tradeoff between repeated edge relaxations and synchronizations.
\item{A heuristic for scheduling the edge relaxations with statistics on the vertex degrees and edge weights.} At the cost of a relatively small number of synchronizations, it guarantees that most of the extended paths are the shortest paths.
\item{A heuristic for reducing the edge traversals with statistics on the vertex degrees and edge weights.} For graphs with a large number of skippable edges, it significantly reduces the traversed edges.
\item{A heuristic algorithm for the SSSP computations on undirected graphs.} Based on the novel method, it exploits the above two heuristics to reduce both the workload and the overhead in parallel settings.
\end{itemize}

The heuristics and the algorithm were evaluated on 73 real-world and synthetic graphs. These graphs span diverse vertex degree distributions and edge weight distributions. On each graph, the synchronization number is comparable to that of Bellman-Ford's algorithm~\cite{22}. On each of the low-diameter graphs, the number of the extended paths is comparable to that of Dijkstra’s algorithm~\cite{23}, the edge traversal number is less than half of the graph's edge number, and the time cost tends to be proportional to the edge traversal number. On 71 graphs, more than 90 percents of the extended paths are the shortest paths. The algorithm was also compared with five state-of-the-art SSSP implementations. On each GAPBS graph except Road, our algorithm's speedup to the best achieved by these five implementations is $2.5\times$ to $5.83\times$.

The rest of this paper is organized as follows: Section 2 analyzes the problem and introduces the novel shortest path search method. Related works are also briefly overviewed in Section 2. The heuristics and the algorithm are described in Section 3. Section 4 presents the evaluation results. And Section 5 concludes the paper.

\section{Problem Analysis and Related Works}
This work considers the SSSP problem on large undirected graphs. It assumes that the vertices are randomly connected and the edges are weighted with random positive reals. Such a graph is represented as a triplet $G=\langle V, E, w\rangle$, where $V$ denotes its vertices, $E$ denotes its edges and $w$ is the weighting function that maps every edge $e\in E$ to a positive real $w(e)$. For convenience, let $\boldsymbol{adj}(u)$ denote the edges incident to vertex $u\in V$ and $\boldsymbol{lsp}(s, u)$ denote the length of the shortest path from $s$ to $u$.

Given a source vertex $s$, the problem is to find the shortest paths from $s$ to other vertices. The result is represented with array \emph{parent[]} and \emph{dist[]}. Initially, $\langle parent[s], dist[s]\rangle$ is $\langle s, 0\rangle$, and $\langle parent[u], dist[u]\rangle$ for every vertex $u\neq s$ is $\langle -1, +\infty\rangle$. The final $\langle parent[u], dist[u]\rangle$ for every $u\in V$ specifies the shortest path linking \emph{s} and $u$. The specified path reaches $u$ via $parent[u]\in V$, and $dist[u]$ is its length $lsp(s, u)$. If \emph{u} is unreachable from \emph{s}, the final $\langle parent[u], dist[u]\rangle$ remains $\langle -1, +\infty\rangle$.  

\subsection{Problem Analysis}\label{Method}
Typically, the SSSP computation is decomposed into concurrent edge relaxations. Each relaxation is executed on some vertex $u\in V$ to relax an adjacent edge $e=(u, v)\in adj(u)$. It \textbf{creates} a new path reaching $v$ via $u$, and updates $\langle parent[v], dist[v]\rangle$ to $\langle u, dist[u]+w(e)\rangle$ if $dist[v]$ is greater than $dist[u]+w(e)$. The path reaching $u$ via $parent[u]$ is referred as the \textbf{extended path}. The relaxation is \emph{redundant} if $dist[u]+w(e)$ is greater than $lsp(s,v)$. The relaxation is also referred to as a \emph{repeated relaxation} if $dist[u]$ is greater than $lsp(s,u)$. If $\langle parent[v], dist[v]\rangle$ has been updated successfully, $v$ is marked as a \emph{frontier}, indicating that the path reaching $v$ via the latest $parent[v]$ is to be extended with the edges of $adj(v)$. The SSSP computation is started by marking $s$ as a frontier, and is completed after the edge relaxations on each frontier has been executed.

To improve the SSSP computation's performance, two issues have to be addressed. The first is to reduce redundant edge relaxations. They not only decrease the SSSP computation's efficiency, but also exacerbate workload imbalance and cause extra communication overhead in parallel settings. The second is to reduce synchronizations. The SSSP computation is realized as rounds of edge relaxations. Relaxations in the same round are synchronized to check the SSSP computation's progress. These synchronizations introduces extra overhead and undermine the efficient utilization of parallel computing resources.

\begin{algorithm}
\floatname{algorithm}{Function}
\footnotesize
\caption{$NLT(lt)$}
\begin{algorithmic}[1]\label{nlt}
	\REQUIRE $G =\langle V, E, w\rangle, s, Frontiers=\{x: x\in V\land lsp(s, x)<lt\}$;
	\FOR{$u \in Frontiers$}
		\STATE{$dist[u] \gets lsp(s, u)$;}	
	  \ENDFOR
	\FOR{$u \in V\textbackslash Frontiers$}
		\STATE{$dist[u] \gets +\infty$;}	
	\ENDFOR
	\STATE{$\langle relaxedVertices, nextFrontiers, returnVal\rangle \gets \langle \{\}, \{\}, +\infty\rangle$;}
	\WHILE{$Frontiers\neq\{\}$}
		\FOR{$u\in Frontiers$}			
			\FOR{$e=(u,v)\in adj(u) \land dist[u] + w(e)\ge lt$}
				\IF{$dist[u]+w(e)<dist[v]$}
					\STATE{$ub\gets dist[v]+\min\{w(x):x\in adj(v)\}$};
					\IF{$v\in relaxedVertices \land returnVal>ub$}
						\STATE{$returnVal \gets ub$};
					\ENDIF
					\STATE{$nextFrontiers\gets nextFrontiers\cup\{v\}$};
					\STATE{$dist[v] \gets dist[u]+w(e)$};
				\ENDIF
			\ENDFOR
		\ENDFOR
		\STATE $relaxedVertices\gets relaxedVertices\cup Frontiers$;
		\STATE $\langle Frontiers, nextFrontiers\rangle \gets \langle nextFrontiers, \{\}\rangle$;
	\ENDWHILE;
    \RETURN{$returnVal$}
\end{algorithmic}
\end{algorithm}

This work proposes an algorithm to study the SSSP computation's redundant relaxations and synchronizations. Function~\ref{nlt} is the algorithm's pseudocode. Given a length threshold $lt\in[0, +\infty)$, the algorithm assumes that a shortest path's length has been known if the length is less than $lt$. By executing edge relaxations to compute the lengths of other shortest paths, it searches a new length threshold $NLT(lt)$, which is the length lower bound of the paths that have been created by repeated relaxations. 

\begin{figure}[!t]
\begin{minipage}[c]{0.48\textwidth}
\centering
\includegraphics[width=0.90\textwidth]{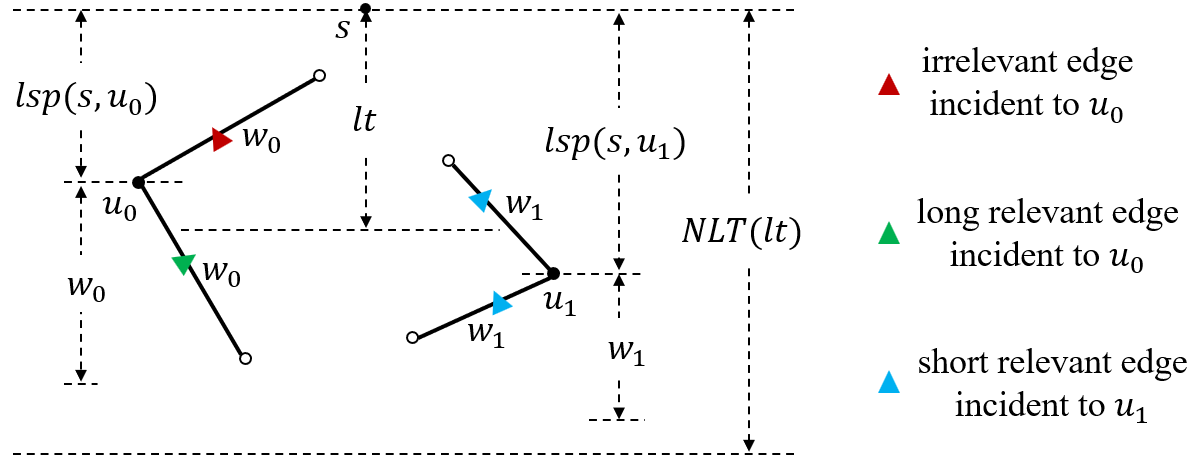}
\end{minipage}
\begin{minipage}[t]{0.48\textwidth}
\centering
\caption{Classification of the edges indexed by $\langle lt, NLT(lt)\rangle$.}
\label{edge-type}
\end{minipage}
\end{figure}

For every vertex whose distance to $s$ is less than $NLT(lt)$, the pair $\langle lt, NLT(lt)\rangle$ is used to index and classify a set of adjacent edges. Edge $e=(u, v)$ with $lsp(s,u)+w(e)\in[lt, NLT(lt))$ is indexed by the pair for vertex $u\in V$. The edge is also indexed by the pair for vertex $v\in V$ if $lsp(s,v)+w(e)$ is within $[lt, NLT(lt))$. As illustrated in Figure~\ref{edge-type}, these indexed edges are classified into \textbf{irrelevant edges}, \textbf{long relevant edges} and \textbf{short relevant edges}. Let $(u, v)\in E$ be an edge indexed by $\langle lt, NLT(lt)\rangle$ for vertex $u$. In the case $lsp(s,u)<lt$ and $lsp(s,v)<lt$, the edge is an irrelevant edge incident to $u$. In the case $lsp(s,u)<lt\le lsp(s,v)$, the edge is a long relevant edge incident to $u$. In the case $lsp(s,u)\ge lt$, the edge is a short relevant edge incident to $u$. 

Shortest paths with lengths within $[lt, NLT(lt))$ are created by relaxations of the indexed short and long relevant edges. For any indexed long relevant edge $(u_0, v_0)$ with $lsp(s,u_0)<lt$, its repeated relaxations on $u_0$ are skipped if the edge has not been relaxed outside Function~\ref{nlt}. For any indexed short relevant edge $e=(u_1, v_1)$ incident to $u_1$, its repeated relaxation on $u_1$ are skipped if paths shorter than $NLT(lt)$ are created with higher priority, since lengths of paths created by these repeated relaxations are at least $NLT(lt)$. 

The above edge indexing and classification mechanism implemented via $\langle lt, NLT(lt)\rangle$ suggests a potentially effective approach for shortest path search. This approach is hereafter called the \textbf{EIC} (\textbf{E}dge \textbf{I}ndexing and \textbf{C}lassification) \textbf{method}. According to Function~\ref{nlt}, there is a finite threshold set for the SSSP computation. Both $0$ and $+\infty$ are in the set. Let $x$ be a threshold in the set. If $x$ is greater than $0$, $y$ with $NLT(y)\equiv x$ is in the set. And $NLT(x)$ is in the set if $x$ is less than $+\infty$. After the shortest paths with the lengths less than $x$ have been created, the shortest paths whose lengths are within $[x, NLT(x))$ can be concurrently created by relaxations of the short and long relevant edges indexed by $\langle x, NLT(x)\rangle$. Furthermore, repeated relaxations can be automatically skipped.

For any length threshold $x\in[0, +\infty)$, it is nontrival to compute $NLT(x)$ and identify the the short and long relevant edges indexed by $\langle x, NLT(x)\rangle$. $NLT(x)$ is computed by creating the shortest paths that are shorter than $NLT(x)$. The indexed edges and their types are also determined by the shortest paths that are shorter than $NLT(x)$. We have proposed two heuristics that exploit statistics of vertex degrees and edge weights to address this issue. One heuristic estimates $NLT(x)$ with these statistics. Another heuristic exploits these statistics to reduce the edge traversed for relaxing the short and long relavant edges. 

\subsection{Related Works}

Over the decades, numerous SSSP algorithms and implementations have been developed. Most of them exploit a vertex-centric approach, i.e., a vertex relaxation is executed on each frontier independently. During the vertex relaxation on $u\in V$, the path that reaches $u$ via the latest $parent[u]$ is extended with every edge of $adj(u)$. The current frontiers are indexed by a worklist, which is initialized with the source vertex $s$. $\forall u\in V$, the vertex is indexed by the worklist after $dist[u]$ has been updated. And $u$ is removed from the worklist after a vertex relaxation has been executed on it. The SSSP computation has been completed if the worklist is empty.

Dijkstra’s algorithm and Bellman-Ford's algorithm are the two classical SSSP algorithms. Both decompose SSSP computations into vertex relaxations. Dijkstra’s algorithm is famous for automatically skipping repeated relaxations. But two vertices can not be relaxed at the same time if their distances to the source vertex are different. Its synchronization overhead is usually proportional to the vertex number, and its resource utilization in parallel settings is usually very poor. Bellman-Ford's algorithm relaxes a vertex $u\in V$ immediately after $dist[u]$ has been updated. It enables the paths with the same number of edges to be extended in parallel, so as to reduce the synchronizations. The cost is that a vertex is repeatedly relaxed if it can be reached via paths with different number of edges.

Many other SSSP algorithms are the variants or combinations of these two classic algorithms. They mainly focus on optimizing the vertex-relaxation order. Some algorithms like $\rho$-stepping algorithm~\cite{19}, Spencer’s algorithm~\cite{24} and DSMR~\cite{25} use a parameter to specify the maximum of vertices that can be relaxed in parallel. Other algorithms like $\Delta$-stepping algorithm~\cite{26}, $\Delta^*$-stepping algorithm~\cite{19}, near-far algorithm~\cite{27} and radius-stepping algorithm~\cite{28} use a parameter to specify the length upper bound for the paths that are extended in parallel. In addition, many techniques have been proposed for scheduling the vertex relaxations efficiently. Lazy batched priority queue~\cite{19}, Fibonacci heap~\cite{29}, smart queue~\cite{30}, ZMSQ~\cite{31}, ordered circular work queue~\cite{32}, bucket fusion~\cite{33} and hybridization~\cite{34} are such techniques.

A few works have attempted to reduce the skippable edges traversed for SSSP computations. For example, Spira’s algorithm~\cite{35} and its variant~\cite{36} create paths in their length order. They introduce a synchronization for every edge relaxation, and assume that the input graph is connected. After every vertex has been visited, the SSSP computation is completed. An edge is skipped if its relaxation creates a path longer than every shortest path. In~\cite{34}, Chakaravarthy et al. utilized the direction optimization technique~\cite{37} to improve $\Delta$-stepping algorithm. A push model and a pull model are available for selecting the long edges relaxed in a step. In the push model, every long edge of \emph{adj(u)} is relaxed on vertex \emph{u} if \emph{u} is a vertex of current bucket. In the pull model, every long edge $(u, v)\in E$ is checked by sending a relaxing request from \emph{u} to \emph{v} if \emph{u} is unsettled; and the request will be responded on \emph{v} by relaxing (\emph{u, v}) if \emph{v} is a vertex of current bucket. When the pull model is chosen by a step, long edges between current bucket and all settled vertices are skipped, at the cost of relaxing requests on each unsettled vertex.

\section{A Heuristic SSSP Algorithm}

The heuristic SSSP algorithm employs the EIC method introduced in~\ref{Method} to solve the SSSP problem on undirected graphs. It dynamically computes a set of scheduling thresholds and selection thresholds. Different scheduling thresholds are computed sequentially in ascending order. The first scheduling threshold is zero. Any two consecutive scheduling thresholds are used to construct a pair for the EIC method. With these pairs, each vertex's adjacent edges are classified into irrelevant edges, long relevant edges and short relevant edges. Every shortest path is created by the relaxation of one long or short relevant edge. The selection thresholds are used to reduce edge traversals for shortest paths that are created by relaxations of the long relevant edges. Each selection threshold is computed for one scheduling threshold pair $\langle lb, ub\rangle$. The selection threshold is at most $lb$ and is denoted $ST(lb, ub)$.

\begin{figure}[!t]
\begin{minipage}[c]{0.48\textwidth}
\centering
\includegraphics[width=0.90\textwidth]{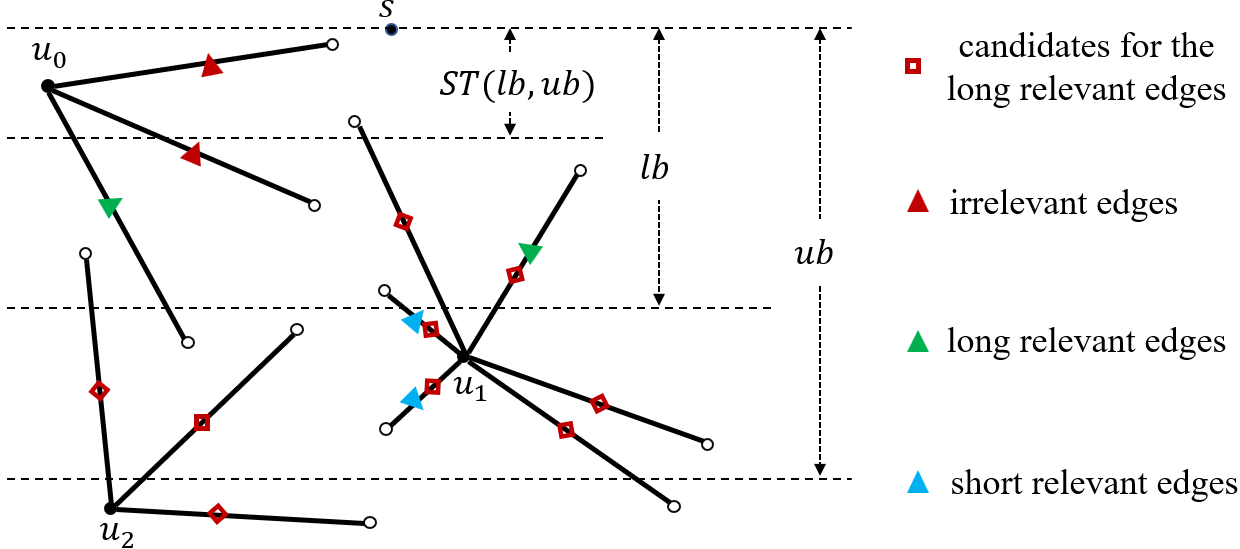}
\end{minipage}
\begin{minipage}[t]{0.48\textwidth}
\centering
\caption{Edges traversed for $\langle lb, ub\rangle$. Each edge traversal starts from a solid-circle vertex and ends at a hollow-circle vertex. The candidate edges are traversed in the pull model. The edges indexed by $\langle lb, ub\rangle$ are traversed in the push model.}
\label{edge-visit}
\end{minipage}
\end{figure}

Given a scheduling threshold pair $\langle lb, ub\rangle$, paths shorter than $lb$ are created before those with lengths at least $lb$. The algorithm takes rounds of edge relaxations to create the shortest paths whose lengths are within $[lb, ub)$. The inner-outer short heuristic ~\cite{34} and the direction optimization technique are employed to select the relaxed edges from those indexed by $\langle lb, ub\rangle$. Paths with different lengths are extended with different models. The traversed edges are illustrated in Figure~\ref{edge-visit}, including the indexed irrelevant edges and long relevant edges incident to each vertex $u_0$ with $lsp(s, u_0)<ST(lb, ub)$, every edge $e_1\in adj(u_1)$ with $w(e_1)<ub-ST(lb, ub)$ and $lsp(s, u_1)\in[lb, ub)$, and each edge $e_2\in adj(u_2)$ with $w(e_2)<ub-ST(lb, ub)$ and $lsp(s, u_2)>lb$. After every shortest path with the length less than $ub$ has been created, the SSSP computation is completed if $ub-\max\{w(x): x\in E\}$ is greater than $\max\{dist[x]: x\in V\land dist[x]<+\infty\}$.

The push model is chosen to extend paths whose lengths are less than $ST(lb, ub)$ or within $[lb, ub)$. The inner-outer short heuristic is employed to ensure that length of the path created by any edge relaxation is within $[lb, ub)$. If the extended path is shorter than $ST(lb, ub)$, every relaxed edge is either an irrelevant edge or a long relevant edge. In the case that the extended path's length is within $[lb, ub)$,  the relaxed edges are short relevant edges. 

The pull mode is chosen to extend the paths whose lengths are within $[ST(lb, ub), lb)$, and the inner-outer short heuristic is employed to reduce candidates of the relaxed edges. On each vertex $u\in V$ with the latest $dist[u]$ greater than $lb$, any adjacent edge $e=(u, v)$ with $w(e)<ub-ST(lb, ub)$ is selected as a candidate edge for those relaxed on vertex $v$, and a request is sent to $v$ for checking whether $e$ is among its long relevant edges indexed by $\langle lb, ub\rangle$. If the latest $dist[v]$ is within $[lb-w(e), ub-w(e))$, $e$ is relaxed when the request arrives at $v$.

The algorithm employs a \textbf{dynamic stepping heuristic} to compute the scheduling thresholds and a \textbf{traversal optimization heuristic} to compute the selection thresholds. Both heuristics are based on the input graph's vertex degree statistics and edge weight statistics. Given a scheduling threshold $lb$ less than $+\infty$, the dynamic stepping heuristic computes a new scheduling threshold $ub$, which is greater than $lb$ and converges to $NLT(lb)$. The traversal optimization heuristic then computes $ST(lb, ub)$.

For any scheduling threshold $lb$, let $\boldsymbol{VS}(lb)$ denote the vertex subset $\{x:x\in V\land lsp(s,x)\ge lb\}$. The following three functions denote the statistics extracted for computing the next scheduling threshold $ub$ and the selection threshold $ST(lb, ub)$. 
\begin{itemize}
\item{$\boldsymbol{sumD}(lb)$:} mapping $lb$ to the total degree of $VS(lb)$.
\item{$\boldsymbol{highD}(lb)$:} mapping $lb$ to a vertex-degree threshold that partitions $VS(lb)$ into two sets. The first set's every vertex has a degree lower than that of any vertex in the second set. The threshold is chosen to minimize the difference in the total vertex degrees between the two sets.  
\item{$\boldsymbol{maxW}(G, ratio):$ mapping a real $ratio\in[0,1]$ to a edge-weight threshold that is at most $\max\{w(x): x\in E\}$. $\forall e\in E$, $ratio$ is the probability of $w(e)\le maxW(G, ratio)$.} 
\end{itemize}

\subsection{Dynamic Stepping}\label{Customization}

The dynamic stepping heuristic assumes that the vertices are randomly connected and the edges are randomly weighted.
Let $lb\in[0, +\infty)$ be such a scheduling threshold that the number of edges between $V\textbackslash VS(lb)$ and $VS(lb)$ is at least $highD(lb)$. Our goal is to construct an appropriate pair $\langle lb, ub\rangle$ for the EIC method. This pair has two characteristics. First, $sumD(ub)$ is comparable to $sumD(lb)/2$ if $highD(lb)$ is greater than the predefined constant $\alpha$. Statistically, most of the short relevant edges indexed by $\langle lb, +\infty\rangle$ are not skippable when $highD(lb)$ is no greater than $\alpha$. Second, paths created by repeated relaxations of Function~\ref{nlt} are highly improbable to be shorter than $ub$.

The dynamic stepping heuristic chooses $lb+gap(lb)$ as $ub$, where $gap(lb)\in(0, maxW(G, 1)]$ is computed with  \eqref{eq-gap} after $VS(lb)$ has been determined. If $highD(lb)$ is at most $\alpha$, $gap(lb)$ is $maxW(G, 1)$. In the case $highD(lb)>\alpha$, $ratio(lb)$ is calculated on the lastest $dist[]$ with \eqref{eq-prob} and  \eqref{eq-ratio}, and $gap(lb)$ is $maxW(G,ratio(lb))$. In \eqref{eq-prob}, $\beta$ is a constant in $(0, 1)$ for tuning the tradeoff between parallelism and efficiency. A small $\beta$ helps reduce repeated relaxations, while a large $\beta$ facilitates creating more shortest paths in parallel. 
\begin{align}
&prob(x)=\min\left(\beta,\textstyle\frac{\max\{sumD(x), 2\times|E|-sumD(x)\}}{2\times|E|}\right)\label{eq-prob}\\
&ratio(x)=1-(1-prob(x))^{1\over{prob(x)\times highD(x)}}\label{eq-ratio}\\
&gap(x)=
\begin{cases}
maxW(G,1)&,\;highD(x)\le\alpha\\
maxW(G,ratio(x))&,\;highD(x)>\alpha
\end{cases}\label{eq-gap}
\end{align}

\textbf{Comparing $sumD(ub)$ and $sumD(lb)/2$.} From \eqref{eq-prob}, we can see $prob(lb)\ge 0.5$ when $\beta$ is in $[0.5, 1)$. In the case $highD(lb)>\alpha$, we show that $sumD(ub)>prob(lb)\times sumD(lb)$ is highly improbable. For convenience, $ub_0\in(ub, +\infty)$ is used to denote the maximal threshold ensuring $sumD(ub_0)\ge prob(lb)\times sumD(lb)$, $wlb$ is used to denote the weight lower bound of the edges between $V\textbackslash VS(lb)$ and $VS(ub_0)$, and $ne$ is used to denote the number of edges between $V\textbackslash VS(lb)$ and $VS(ub_0)$.

We know $wlb\ge ub_0-lb$. At the same time, the probability of $wlb>maxW(G, ratio(lb))$ is $(1-ratio(lb))^{ne}$. Therefore, the probability of $ub_0-lb>maxW(G, ratio(lb))$ is at most $(1-ratio(lb))^{ne}$. Furthermore, $gap(lb)=ub-lb$ is $maxW(G, ratio(lb))$, since $\alpha$ is greater than $highD(lb)$. In turn, the probability of $ub_0>ub$ is at most $(1-ratio(lb))^{ne}$. In other words, the probability of $sumD(ub)>prob(lb)\times sumD(lb)$ is at most $(1-ratio(lb))^{ne}$, which is  $(1-prob(lb))^{{ne}\over{prob(lb)\times highD(lb)}}$. As the vertices are randomly connected, $ne$ is statistically about the product of $prob(lb)$ and the number of edges between $V\textbackslash VS(lb)$ and $VS(lb)$. Therefore, this probability is less than $1-prob(lb)$ and quickly decreases with the number of edges between $V\textbackslash VS(lb)$ and $VS(lb)$.

\textbf{Repeated relaxations}. For any vertex $u\in V$ with $lsp(s, u)\in[lb, ub)$, let $llb$ be the length lower bound of paths created in Function~\ref{nlt} by repeated relaxations executed on $u$. We show that $llb<ub$ is highly improbable. In the case $highD(lb)\le\alpha$, repeated relaxations are highly unlikely to be executed in Function~\ref{nlt}. In the case $dist[u]$ is finalized by the first round of relaxations, repeated relaxations are impossible to be executed on $u$. We consider the case $highD(lb)>\alpha$ and $dist[u]$ is finalized after the first round of relaxations. 

Let $wlb$ denote $\min\{w(x):x\in adj(u)\}$, we know $llb-wlb>lsp(s,u)$. Since $dist[u]$ is finalized after the first round of relaxations, $lb+wlb$ is at most $lsp(s, u)$. In turn, $llb$ is greater than $lb + 2\times wlb$. At the same time, the probability of $wlb>gap(lb)$ is $(1-ratio(lb))^{|adj(u)|}$, since the probability of $w(e)>gap(lb)$ for any $e\in adj(u)$ is $1-ratio(lb)$. The probability of $2\times wlb>gap(lb)$ is then greater than $(1-ratio(lb))^{|adj(u)|}$. And the probability of $llb>lb+2\times wlb>lb+gap(lb)=ub$ is also greater than $(1-ratio(lb))^{|adj(u)|}$. If $|adj(u)|$ is less than $highD(lb)$, $(1-ratio(lb))^{|adj(u)|}=(1-prob(lb))^{{|adj(u)|}\over{(prob(lb)\times highD(lb))}}$ is greater than $1-prob(lb)$ and increases quickly as $|adj(u)|$ decreases. As the degrees of most vertices in $VS(lb)$ are no greater than $highD(lb)$, it is highly improbable for $llb$ to be less than $ub$.

\subsection{Traversal Optimization}\label{Selection}

The traversal optimization heuristic aims to reduce edge traversals for shortest paths that are created by relaxations of long relevant edges. Let $\langle lb, ub\rangle$ be a pair constructed for the EIC method. Based on the vertex-degree statistics, the traversal optimization heuristic chooses a selection threshold $ST(lb, ub)$ from $[0, lb]$. Given a vertex $u\in V$ with $lsp(s, u)<ST(lb, ub)$, the shortest path linking $s$ and $u$ is extended with the direction optimization technique's push model, and each of its adjacent edges indexed by $\langle lb, ub\rangle$ is relaxed. For each vertex $u\in V$ with $lsp(s, u)\in[ST(lb, ub), lb)$, the direction optimization technique's pull model is used to extend the shortest path between $s$ and $u$, so as to skip relaxations of the irrelevant edges indexed by $\langle lb, ub\rangle$.

We analyze which threshold in $[0, lb]$ is most appropriate for $ST(lb, ub)$. For any $x\in[0, lb)$, the following three functions are introduced.
\begin{itemize}
\item{$pushed(x, lb, ub)$: the cardinality of a multiset comprising all edges that are indexed by $\langle lb, ub\rangle$ and incident to vertices of $VS(x)\textbackslash VS(lb)$.} 
\item{$long(x, lb, ub)$: the number of long relevant edges that are indexed by $\langle lb, ub\rangle$ and incident to vertices of $VS(x)\textbackslash VS(lb)$.}
\item{$pulled(x, lb, ub)$: the cardinality of a multiset comprising all edges that are incident to vertices of $VS(lb)$ and their weights are less than $ub-x$.}
\end{itemize}

When any $x\in[0, lb)$ is chosen as $ST(lb, ub)$, $profit(x, lb, ub)=pushed(x, lb, ub)-long(x, lb, ub)-pulled(x, lb, ub)$ is the number of edge traversals reduced with the traversal optimization heuristic. Let $lb_0\in[0, lb)$ be such a threshold that $profit(lb_0, lb, ub)$ is no less than $profit(x, lb, ub)$ for any $x\in[0, lb)$. If $profit(lb_0, lb, ub)$ is positive, $lb_0$ is chosen as $ST(lb, ub)$; otherwise, $lb$ is chosen.

\begin{align}
&pushed(x, lb, y)=\textstyle{{{(y-lb)\times(sumD(x)-sumD(lb))}\over{maxW(G,1)}}}&\label{eq-pushed}\\
&pulled(x, lb, y)=(y-x)\times sumD(lb)/maxW(G,1)&\label{eq-pulled}\\
&long(x, lb, y)=pulled(x, lb, y)\times\textstyle{{sumD(x)-sumD(lb)}\over{2\times|E|}}&\label{eq-long}
\end{align}

The traversal optimization heuristic assumes that the vertices are randomly connected and the edges are randomly weighted. Given a threshold $x\in[0, lb)$, it leverages vertex-degree statistics to estimate $profit(x, lb, ub)$. Let $lb_0$ denote $\max\{x, lb-maxW(G,1)\}$, $ub_0$ denote $\min\{ub, lb+maxW(G,1)\}$, and $ub_1$ denote $\min\{ub, lb_0+maxW(G,1)\}$. If a path with the length in $[x, lb)$ is extended with edges indexed by $\langle lb, ub\rangle$, the path's length is at least $lb_0$, lengths of the new paths are at most $ub_0$, and every relaxed edge's weight is at most $ub_1-lb_0$. Therefore, the value of $pushed(x, lb, ub)$ is identical to that of $pushed(lb_0, lb, ub_0)$, which is estimated with \eqref{eq-pushed}. The value of $pulled(x, lb, ub)$ is identical to that of $pulled(lb_0, lb, ub_1)$, which is estimated with \eqref{eq-pulled}. The value of $long(x, lb, ub)$ is identical to that of $long(lb_0, lb, ub_1)$, which is estimated with \eqref{eq-long}.  

\subsection{The Control Flow}\label{ctl-overhead}
On the input graph $G=\langle V, E, w\rangle$, the heuristic SSSP algorithm takes a sequence of steps to update $\langle parent[], dist[]\rangle$. Each step is specified with a scheduling threshold pair constructed for the EIC method, and executes rounds of edge relaxations. The control flow is realized with $\langle lb, ub\rangle$, $st$ and \emph{Frontiers}.
\begin{itemize}
\item{$\langle lb, ub\rangle$: the current step's scheduling threshold pair. } 
\item{\emph{st}: the selection threshold chosen for $\langle lb, ub\rangle$.}
\item{\emph{Frontiers}: the vertex set indexing the paths that are extended by the next round of edge relaxations}
\end{itemize}

The current step creates the shortest paths whose lengths are within $[lb,ub)$. The edge relaxations are selected with $\langle lb, ub, st\rangle$ and are scheduled with $Frontiers$. Both $st$ and $ub$ are initialized before the current step's edge relaxations, and $Frontiers$ is initialized after the first round of edge relaxations. The current step's $ub$ is the next step's $lb$. After the current step, the SSSP computation is completed if the latest $ub-maxW(G,1)$ is greater than $\max\{dist[x]: x\in V\land dist[x]<+\infty\}$.

The first step's initial $\langle lb, ub\rangle$ is $\langle 0, +\infty\rangle$. After each round of edge relaxations, $ub$ is updated to the length lower bound of paths that link $s$ and vertices with degrees at least $highD(0)$. It is expected that $ub$ is finalized before a large number of edge relaxations have been executed. 

Due to two issues, the flow-controlling overhead is non-ignorable. First, relaxations in the same round are synchronized. Our solution is to defer the relaxation of long relevant edges in the push model until after the first round of edge relaxation. These deferred relaxations share synchronizations with those of the short relevant edges. Second, the input graph's vertex degree statistics and edge weight statistics are referred in each step. If $sumD(lb)$ is at least $|E|$, the algorithm chooses $lb$ as $st$, so as to reduce the statistics extraction overhead.

\begin{algorithm}
\renewcommand{\algorithmicrequire}{\textbf{Input:}}
\renewcommand{\algorithmicensure}{\textbf{Output:}}
\footnotesize
\caption{The heuristic SSSP algorithm}
\begin{algorithmic}[1]\label{alg0}
	\REQUIRE $G =\langle V, E, w\rangle, s$;
	\ENSURE $dist[], parent[]$;
	\FORALL{$u \in V$} \STATE{$\langle dist[u], parent[u] \rangle \gets \langle +\infty, -1\rangle$;} \ENDFOR
	\STATE $\langle dist[s], parent[s], st, lb, ub\rangle\gets\langle0, s, 0, 0, +\infty\rangle$;
	\REPEAT
		\STATE $Frontiers\gets initFrontiers(st, lb, ub)$;
		\WHILE{$Frontiers \neq \{\}$}
			\STATE $Paths\gets Frontiers-\{x:x\in V\land dist[x]>0\land|adj(x)|\equiv1\}$;
			\STATE $Frontiers\gets \{\}$;
			\FORALL{$u \in Paths$}
				\STATE{$selected\gets \{x: x\in adj(u)\land lb \le dist[u] + w(x) < ub\}$};
				\FORALL{$e = (u, v)\in selected \land v\neq parent[u]$} 
		  		  \IF{$CAS(\langle dist[v], parent[v]\rangle , \langle dist[u] + w(e), u\rangle )$}\STATE $SAP (Frontiers, v)$;\ENDIF
				\ENDFOR
			\ENDFOR
			\IF{$lb\equiv 0$}
				\STATE{$ub\gets\min\{dist[x]: x\in V\wedge|adj(x)|\ge highD(lb)\}$};
    			\ENDIF
		\ENDWHILE
		\STATE $\langle st, lb,ub\rangle\gets\langle computeST(lb, ub), ub,ub+gap(ub)\rangle$;
	\UNTIL {$\max\{dist[x]: x\in V\land dist[x]<+\infty\}+maxW(G,1)<lb$};
    \RETURN
\end{algorithmic}
\end{algorithm}

\begin{algorithm}
\renewcommand{\thealgorithm}{1}
\floatname{algorithm}{Function}
\footnotesize
\caption{$initFrontiers(st, lb, ub)$}
\label{fun1}
\begin{algorithmic}[1]
\STATE $lb_0\gets\max\{0,lb-maxW(G, 1)\}$;
\STATE $Frontiers\gets\{x: x\in V\land lb_0\le dist[x]\le st\}$;
\IF{$st\equiv lb$}
	\RETURN{$Frontiers\cup \{x:x\in V\land lb\le dist[x]<ub\}$};
\ENDIF
\FORALL{$u \in\{x:x\in V\land dist[x]>lb\}$}
	\STATE $selected\gets \{x: x\in adj(u)\land w(x) < ub-st\}$;
	\FORALL{$e = (u, v) \in selected$} 
		\IF{$st\le dist[v]<lb\;\AND\;dist[v] + w(e)< ub$}
			\STATE $CAS(\langle dist[u], parent[u]\rangle, \langle dist[v] + w(e), v\rangle)$;	
		\ENDIF
	\ENDFOR
\ENDFOR
\RETURN{$Frontiers\cup \{x: x\in V\land lb\le dist[x]<ub\}$};
\end{algorithmic} 
\end{algorithm}

\begin{algorithm}
\renewcommand{\thealgorithm}{2}
\floatname{algorithm}{Function}
\footnotesize
\caption{$computeST(lb, ub)$}
\label{fun2}
\begin{algorithmic}[1]
	\STATE $\langle st, maxProfit\rangle\gets\langle ub,0\rangle$;
	\IF{$sumD(ub)\ge|E|\lor gap(lb)\equiv maxW(G, 1)$}
		\RETURN{$st$};
	\ENDIF
	
	\IF{$gap(ub)\equiv maxW(G, 1)$}
			\RETURN{$ ub-maxW(G, 1)$}; 
	\ENDIF
	\STATE $y\gets\ ub+gap(ub)\rangle$;
	\FOR{$x\in \{dist[y]:y\in V\}\land x<ub$}					
		\STATE $profit\gets\ pushed(x,ub,y)-long(x,ub,y)-pulled(x,ub,y)$;
		\IF{$maxProfit<profit$}
			\STATE $\langle st,maxProfit\rangle\gets\langle x,profit\rangle$;
		\ENDIF
	\ENDFOR
\RETURN{$st$};
\end{algorithmic} 
\end{algorithm}

Algorithm~\ref{alg0} is the algorithm's pseudo code. Function~\ref{fun1} identifies and executes the current step's first round of edge relaxations. It also initializes \emph{Frontiers}. Then the successive rounds of edge relaxations are identified and executed with \emph{Frontiers}. If the current step is the first step, the algorithm attempts to update $ub$ after each round of edge relaxations. After each step, Function~\ref{fun2} chooses an appropriate $st$ for the next step, and $lb$ is replaced with the latest $ub$.

Two atomic operations are used to address the collisions caused by parallel edge relaxations. One is the compare-and-swap operation $\boldsymbol{CAS}(\langle x_0, y_0\rangle, \langle x_1, y_1\rangle)$ that updates $\langle x_0, y_0\rangle\;\text{to}\;\langle x_1, y_1\rangle$ if $x_0$ is greater than $x_1$. If  $x_0$ has been successfully updated, the operation returns $\boldsymbol{true}$; otherwise, it returns $\boldsymbol{false}$. Another is the search-and-push operation $\boldsymbol{SAP}(vset, v)$ that pushes vertex $v$ into vertex set $vset$ if $v$ is not found in $vset$. 

\subsection{Efficiency Analysis}\label{efficiency-analysis}

Based on the dynamic stepping and the traversal optimization, the heuristic SSSP algorithm solves SSSP problems with the EIC method. It constructs a set of scheduling threshold pairs for the SSSP computation, and selects a selection threshold for each pair. With these pairs, each vertex's adjacent edges are classified into irrelevant edges, long relevant edges and short relevant edges. The scheduling threshold pairs and the selection thresholds are used to decompose the SSSP computation into concurrent edge traversals. An edge is traversed when it is relaxed or an edge-relaxing request is sent along it. The edge traversals are scheduled with these pairs, allowing edge indexed by the same pair to be relaxed in parallel. 

By heuristically constructing the scheduling threshold pairs, the dynamic stepping helps the EIC method to significantly reduce repeated relaxations at the cost of almost negligible synchronizations. In addition, the traversal optimization helps the EIC method to reduce edge traversals for shortest paths that are created by relaxations of the long relevant edges.


\begin{figure}[!t]
\begin{minipage}[c]{0.48\textwidth}
\centering
\includegraphics[width=0.90\textwidth]{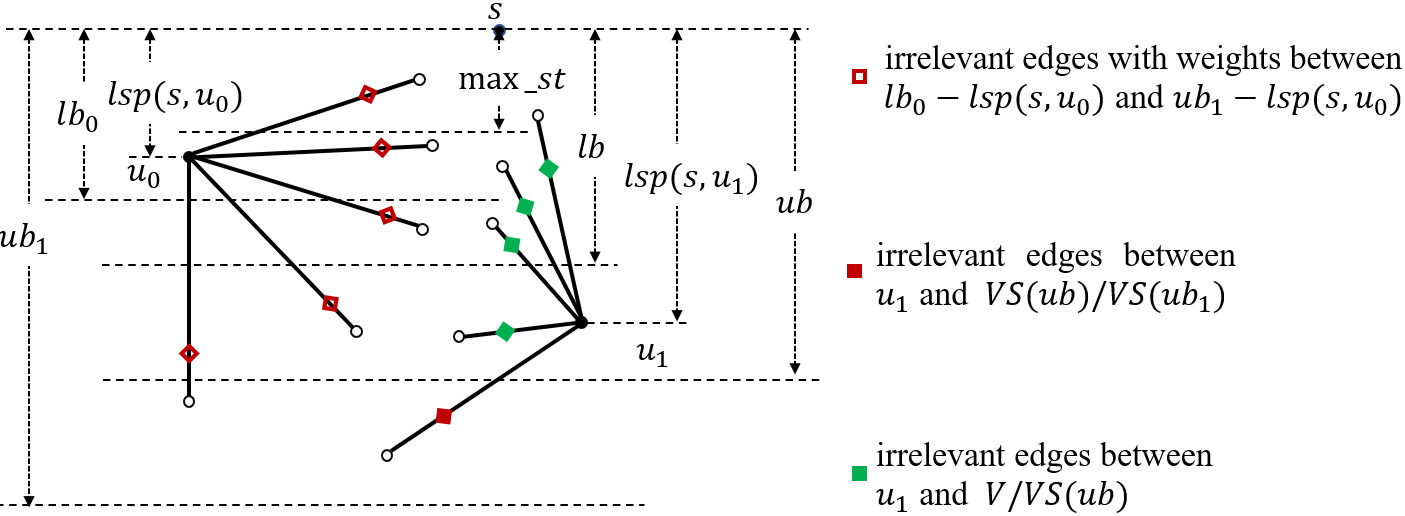}
\end{minipage}
\begin{minipage}[t]{0.48\textwidth}
\centering
\caption{Skipped irrelevant edges indexed by $\langle lb, ub\rangle$. Each skipped traversal starts from a solid-circle vertex.  The work in~\cite{34} only supports skipping the irrelevant edges between $V\textbackslash VS(ub)$ and each vertex $u_1$ with $lsp(s, u_1)\in [lb, ub)$.}
\label{edge-skipped}
\end{minipage}
\end{figure}

When $|E|$ is no less than a few times of $|V|$, $highD(0)$ is greater than $\alpha$, and the algorithm automatically skips a substantial fraction of irrelevant edges. Figure~\ref{edge-skipped} illustrates the skipped edges. Let $\langle lb_0, ub_0\rangle$ be the first pair with $ST(lb_0, ub_0)<lb_0$, $\langle lb_1, ub_1\rangle$ be the last pair with $ub_1-maxW(G,1)<ST(lb_1, ub_1)<lb_1$, and $max\_st$ be the selection threshold that is no less than $ST(x, y)$ for any pair $\langle x, y\rangle$ with $maxW(G, 1)> y-ST(x,y)$. Given a pair $\langle lb, ub\rangle$ with $lb\ge lb_0$ and $ub\le ub_1$, the following three types of edges are skipped.
\begin{itemize}
\item{Each edge $e_0\in adj(u_0)$ with $lsp(s, u_0)+w(e_0)\in [lb_0, ub_1)$ and $lsp(s, u_0)<lb_0$: for vertex $u_0$, $e_0$ is an irrelevant edge indexed by some constructed pair $\langle x, y\rangle$ with $x\in[lb_0, lb_1]$.} 
\item{Each edge $e_2=(u_1, v_2)$ with $lsp(s, v_2)>ub$ and $lsp(s, u_1)+w(e)\in [x, y)$: for vertex $u_1$, $e_1$ is an irrelevant edge indexed by the constructed pair $\langle x, y\rangle$ with $x\in[ub, lb_1]$.}
\item{Each edge $e_1=(u_1, v_1)$ with $lsp(s, v_1)<ub$ and $lb<lsp(s, u_1)<ub\le lsp(s, u_1)+w(e)$: for vertex $u_1$, $e_1$ is an irrelevant edge indexed by some constructed pair $\langle x, y\rangle$ with $x\in[lb, lb_1]$.}
\end{itemize}

\section{Evaluation}
The dynamic stepping heuristic, the traversal optimization heuristic and the heuristic SSSP algorithm were evaluated on 73 graphs, including GAPBS benchmark graphs, Graph500 benchmark graphs and their variants. The experiments were conducted on a Supermicro SYS-2049U-TR4 server running CentOS Linux release 7.8.2003. There are 4 Intel(R) Xeon(R) Gold 5218 CPUs @ 2.30GHz and 1536 GB of memory, each CPU has 16 cores. In the rest, the algorithm's implementation is referred to as \emph{EIC}. 

For each graph, 64 vertices with nonzero degrees were randomly selected, and each vertex served as the source vertex of one trial. Different trials were run one after another. The reported results are the average of 64 trials. Besides time cost, the following symbols are used to report the experimental results. The reported results are normalized, so that they are comparable between different graphs. 
\begin{itemize}
\item{\emph{nFrontier}: the number of paths that have been extended. The reported result is normalized by the number of non-leaf vertices that are reachable from the source vertex. }
\item{\emph{nSync}: the number of synchronizations. The reported result is normalized by the vertex number's logarithm.}
\item{\emph{nTrav}: the computation's number of edge traversals. The reported result is normalized by the number of vertices that are reachable from the source vertex.} 
\end{itemize}

\subsection{Implementation}
EIC was developed with C and MPICH 4.1.2~\cite{38}. It implements the heuristic SSSP algorithm as parallel MPI (Message-Passing Interface) processes. Different processes exchange messages asynchronously. The input graph's vertices are distributed evenly among the processes. Every vertex is owned by one process, and its incident edges are also stored in the owner process. Every vertex is preprocessed by sorting the incident edges in the weight order. In addition, the graph's edge weights are preprocessed to quantize their distribution. The quantizing results are saved in an array \emph{RtoW[]}, whose size is \emph{RATIO\_NUM}. $\forall x\in[0,RATIO\_NUM)$, \emph{RtoW[x]} is $maxW(G,\frac{x}{RATIO\_NUM-1})$.

$\forall e=(u, v)\in E$, let $proc_u$ denote $u$'s owner process and $proc_v$ denote $v$'s owner process. If $proc_u$ is different from $proc_v$, the two processes cooperate asynchronously via MPI messages to execute relaxations on $e$ and send edge-relaxing requests along $e$.
\begin{itemize}
\item{Case A: $e$ is executed on to create the path that reaches \emph{v} via \emph{u}. In $proc_u$, $\langle RELAX, u, v, dist[u]+w(e)\rangle$ is pushed into some message that will be sent to $proc_v$. When the message arrives at $proc_v$, the latest $dist[v]$ is compared with $dist[u]+w(e)$, and the pair $\langle parent[v], dist[v]\rangle$ is updated to $\langle u, dist[u]+w(e)\rangle$ if $dist[v]$ is greater than $dist[u]+w(e)$.} 
\item{Case B: an edge-relaxing request is sent from $u$ to $v$. In $proc_u$, $\langle REQUEST, u, v, w(e)\rangle$ is pushed into some message that will be sent to $proc_v$. When the message arrives at $proc_v$, both $dist[v]$ and $dist[v]+w(e)$ are checked with $\langle st, lb, ub\rangle$. If $dist[v]$ is within [\emph{st}, \emph{lb}) and $dist[v]+w(e)$ is less than \emph{ub}, $proc_v$ executes an relaxation on $v$ to relax $e$.}
\end{itemize}

Both $\langle ub, st\rangle$ and \emph{Frontiers} require different MPI processes to cooperate synchronously. EIC distinguishes the following two cases to reduce the flow-controlling overhead, where \emph{ST\_NUM} and \emph{FUSED} are predefined integer constants.
\begin{itemize}
\item{$sumD(lb)\le|E|\land gap(lb)<maxW(G,1)$. Let $st_0$ denote $\min\{lsp(s,x):x\in V\land |adj(x)|\ge highD(0)\}$ and $st_1$ denote the first scheduling threshold calculated by EIC that $sumD(st_1)$ is at most \emph{|E|}. EIC reduces the statistic-extracting overhead by choosing \emph{sw} from $\{x\times\frac{st_1-st_0}{ST\_NUM}+st_0:0\le x\le ST\_NUM\}$. }
\item{$gap(lb)\equiv maxW(G,1)$. EIC exploits the bucket fusion~\cite{33} to execute the first \emph{FUSED} rounds of edge relaxations concurrently, so as to improve load balance and reduce synchronizations. Specifically, each MPI process loops \emph{FUSED} iterations on its local \emph{Frontiers} before a synchronization. Each iteration executes three types of operations: (1) creating the paths specified by current local \emph{Frontiers}, (2) processing the messages that have arrived, and (3) posting the requests for sending the messages that have been created.}
\end{itemize}

In the experiments, the heuristic SSSP algorithm was configured with $\langle\alpha, \beta\rangle=\langle3, 0.9\rangle$. The reported results were achieved with $\langle RATIO\_NUM, ST\_NUM, FUSED\rangle=\langle 2^{12}, 2^{10}, 2^{8}\rangle$. 

\subsection{Datasets}
The datasets consist of 9 benchmark graphs and 64 variant graphs. The benchmark graphs are listed in Table~\ref{tab-graphs}. Every benchmark graph prefixed with \emph{gr26\_} is a synthetic graph that was created by the graph generator in Graph500 3.0.0. The synthetic graph's edge list is specified with the argument pair $\langle scale, edge\_factor\rangle$, where \emph{scale} is 26 and \emph{edge\_factor} is the graph name's suffix number. Given a synthetic graph specified with $\langle scale, edge\_factor\rangle$, its vertex number is $2^{scale}$, its edge number is $edge\_factor\times2^{scale}$, and its edge weights are uniformly distributed in (0,1]. Every synthetic graph's isolated vertices are not counted by the $|V|$ column of Table~\ref{tab-graphs}. In Table~\ref{tab-graphs}, a graph is from the GAPBS benchmark if its name is not prefixed with \emph{gr26\_}. Every GAPBS benchmark graph is a real-world network. It is notable that the directed benchmark graphs are symmetrized to be undirected in our experiments.

Every synthetic graph in Table~\ref{tab-graphs} was used to create 16 variant graphs. Each variant graph exploited an integer \emph{power} or a float \emph{pivot} to modify its original benchmark graph's edge weights. Integer \emph{power} is selected from \{1, 2, 3, 4, 6, 8, 10\}. Float \emph{pivot} is selected from\{0.1, 0.2, 0.3, 0.4, 0.5, 0.6, 0.7, 0.8, 0.9\}.
\begin{itemize}
\item{A variant graph specified with the integer \emph{power}: the benchmark graph's edge weights are mapped to $2^{power}-1$ integers. $\forall e\in E$, the benchmark graph's weight \emph{w(e)} is replaced by $discretize(w(e),power)$, an integer calculated with \eqref{eq-discete}.}
\item{A variant graph specified with the float \emph{pivot}: $\forall e\in E$, the benchmark graph's weight \emph{w(e)} is replaced by the float $converge(w(e),pivot)$ that is calculated with \eqref{eq-weight}. The new weight distribution is a bell curve whose peak is at \emph{pivot}, and half of the new weights are less than \emph{pivot}.}
\end{itemize}

\begin{align}
&discretize(x,y)=1+x\times(2^y-2)\label{eq-discete}\\
&converge(x,y)=
\begin{cases}
y-y\times(1-2x)^2,&x\le0.5\\
y+(1-y)\times(1-2x)^2,&x>0.5
\end{cases}\label{eq-weight}
\end{align}
\begin{table}[!t]
\footnotesize
\caption{Original benchmark graphs}
\label{tab-graphs}
\tabcolsep 5.0pt 
\begin{tabular*}{0.48\textwidth}{l|rrrrr}
\toprule
        & $|E|$ & $|V|$ & $\max\limits_{x\in V}|adj(x)|$ & highD(0) & ${{|E|-|V|}\over{|E|}}$\\\hline
gr26\_4 & 268.4(M)&21.7(M)&426944&417&0.919\\
gr26\_8	& 536.8(M)&27.1(M)&854253&837&0.950\\
gr26\_16&1073.7(M)&32.8(M)&1709303&1681&0.969\\
gr26\_32&2147.5(M)&38.6(M)&3417833&3370&0.982\\
Road&29.2(M)&23.9(M)&9&3&0.179\\
Urand&2147.5(M)&134.2(M)&68&33&0.937\\
Web&1930.3(M)&50.6(M)&8563852&176&0.974\\
Twitter&1468.4(M)&41.7(M)&3081112&866&0.972\\
Kron&2147.5(M)&63.1(M)&2598214&2517&0.971\\
\bottomrule
\end{tabular*}
\end{table}

\subsection{Evaluation Results of the Dynamic Stepping}

The dynamic stepping helps the EIC method to significantly reduce repeated relaxations at the cost of almost negligible synchronizations. We use $nFrontier$ to measure the repeated relaxations that have been executed. If none of repeated relaxations has been executed, every extended path is a shortest path, and $nFrontier$ is 1. The synchronizations introduced by the EIC method are measured with $nSync$, which relies on the number of scheduling threshold pair that the heuristic has constructed for the SSSP computation. Less is $nSync$, fewer synchronizations have been introduced, and more shortest paths have been created in parallel.
 
\begin{figure}[!t]
\begin{minipage}[c]{0.48\textwidth}
\centering
\subfloat[]{\includegraphics[width=0.45\textwidth]{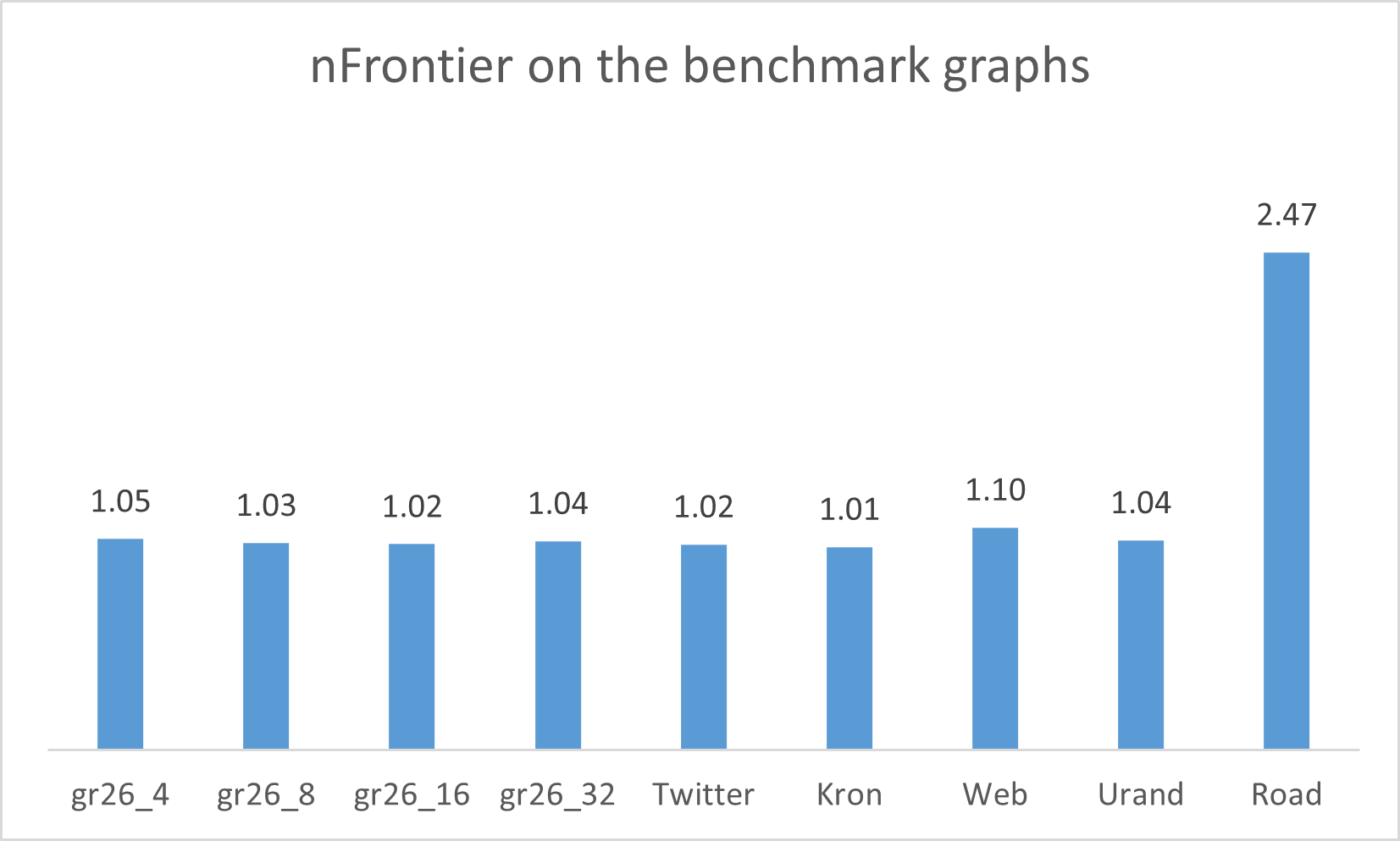}
\label{nFrontier-benchmark}}
\hfil
\subfloat[]{\includegraphics[width=0.45\textwidth]{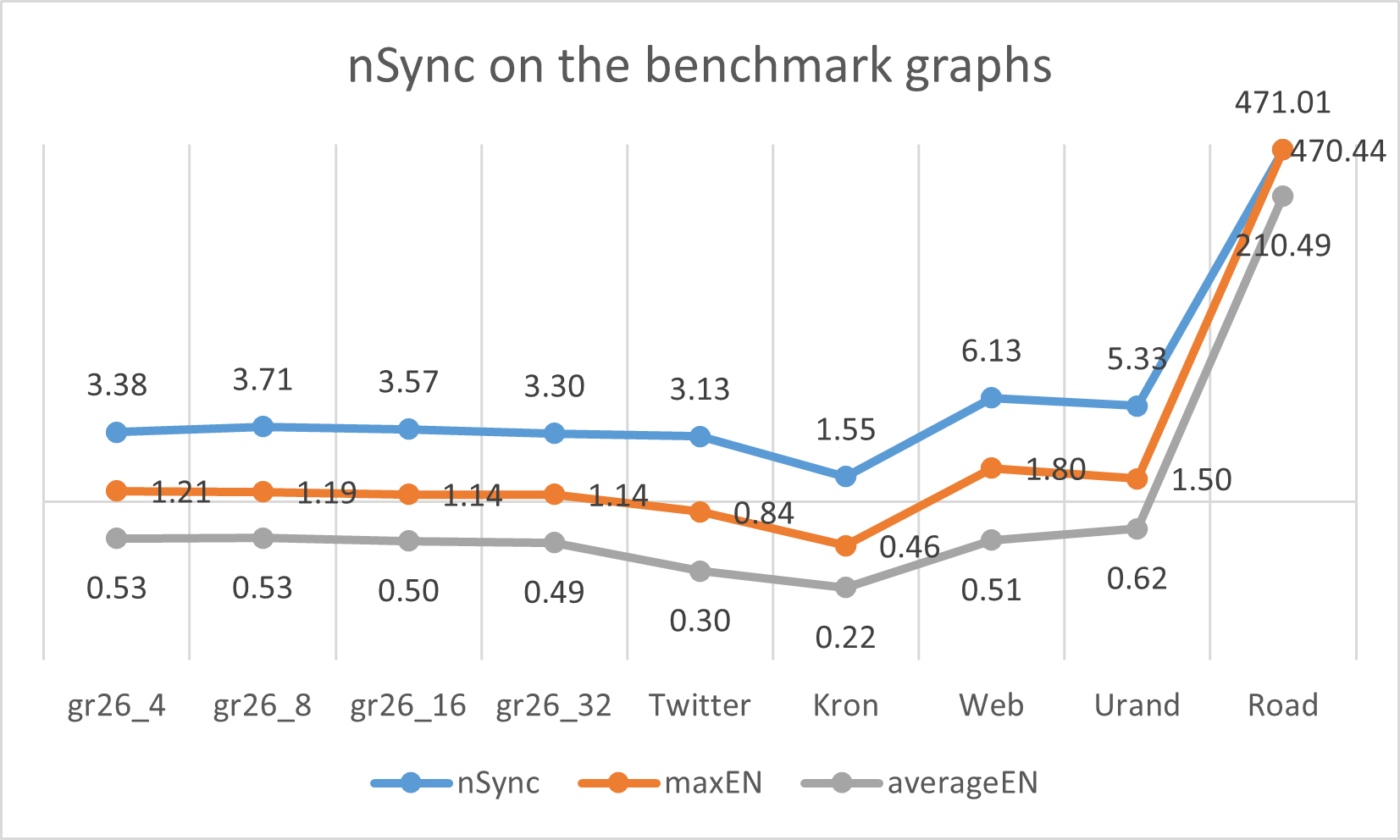}
\label{nSync-benchmark}}
\end{minipage}
\begin{minipage}[c]{0.48\textwidth}
\centering
\subfloat[]{\includegraphics[width=0.45\textwidth]{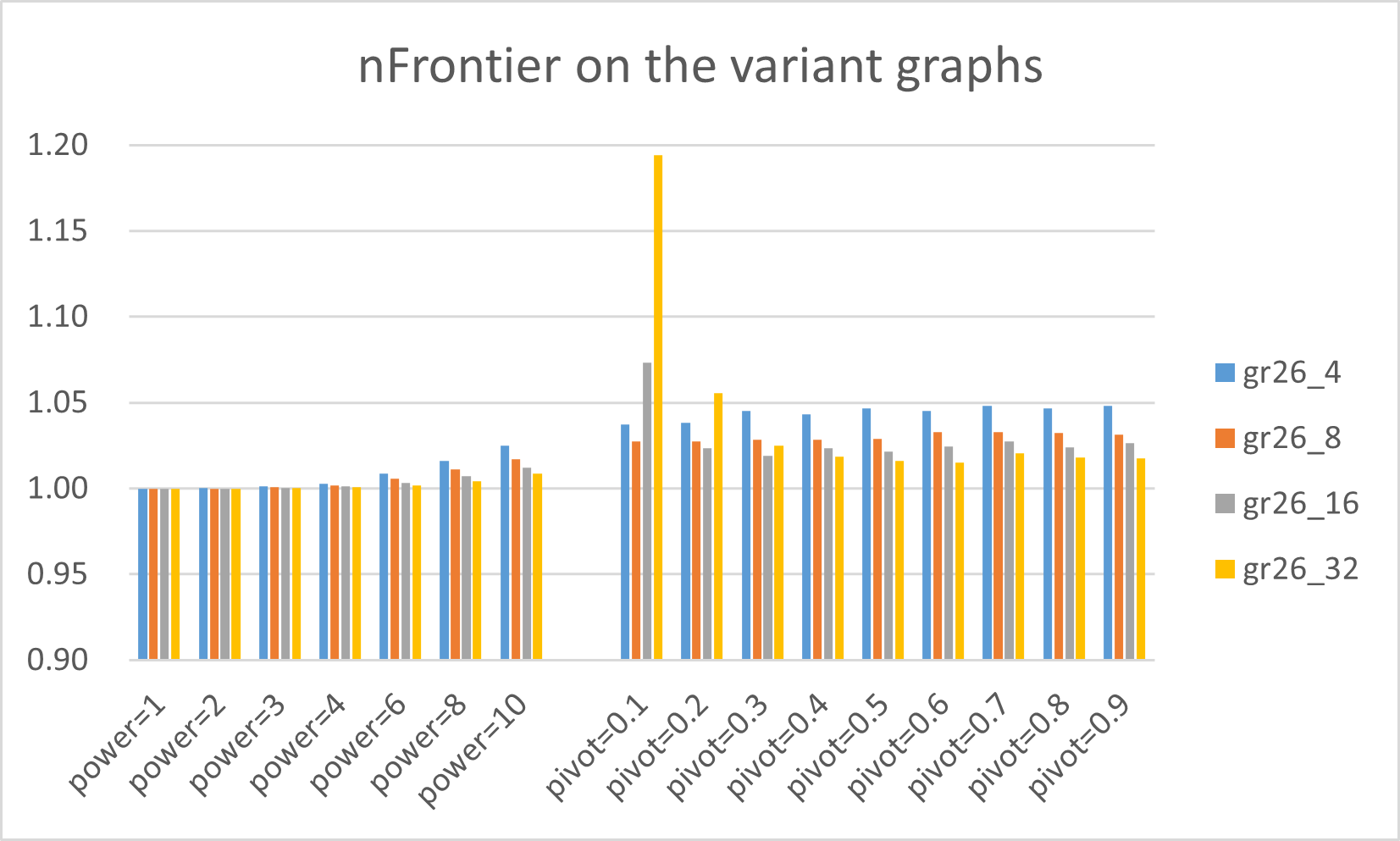}
\label{nFrontier-variant}}
\hfil
\subfloat[]{\includegraphics[width=0.45\textwidth]{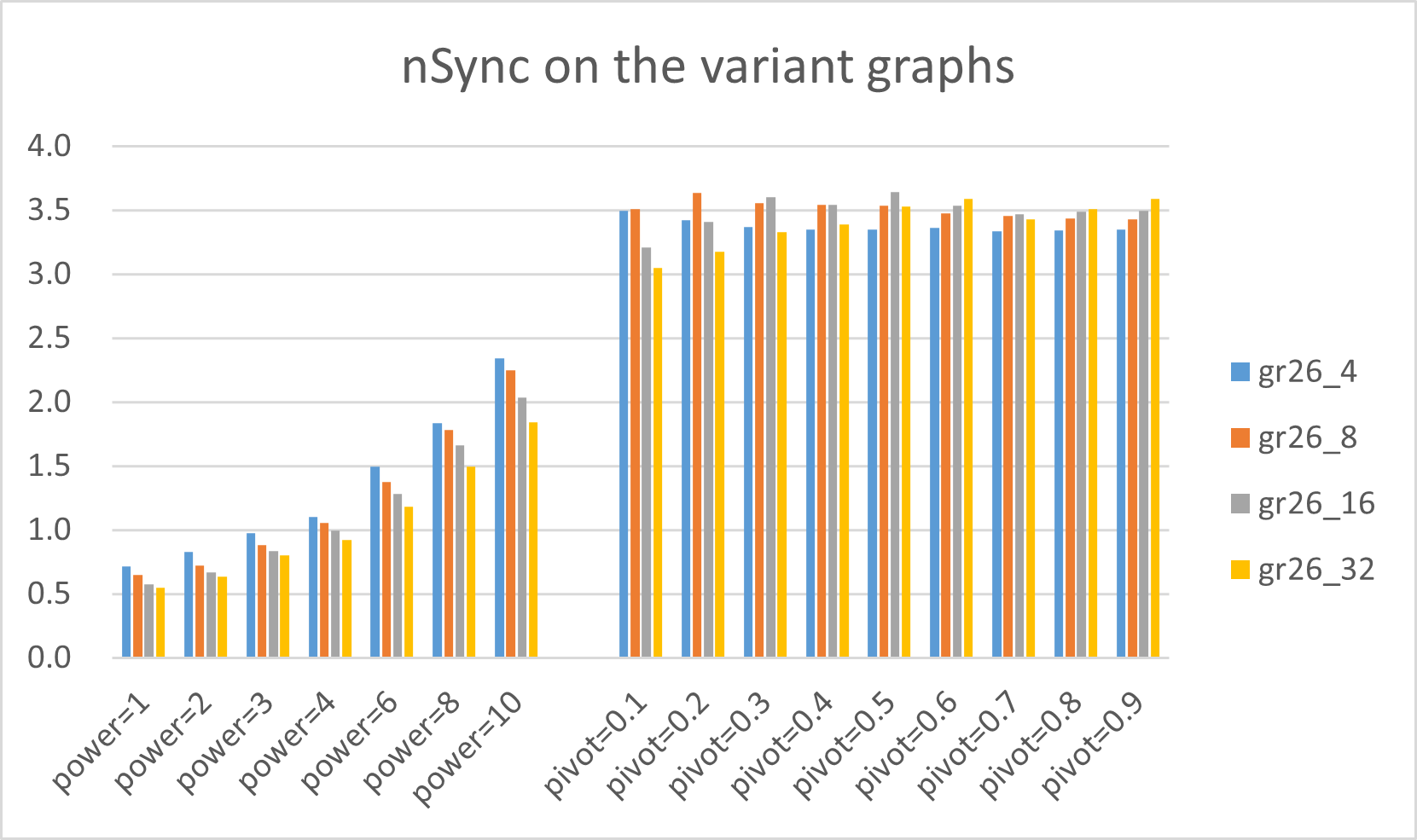}
\label{nSync-variant}}
\end{minipage}
\begin{minipage}[t]{0.48\textwidth}
\centering
\caption{Evaluation results: nFrontier and nSync.} 
\label{customization}
\end{minipage}
\end{figure}

On each benchmark graph $\langle V, E, w\rangle$, the achieved $nSync$ is compared with two statistics that reveal the runtime's dominant factor. Both are normalized by $\log_2{(|V|)}$. One is the average edge number of the shortest paths, denoted as $averageEN$. Another is the maximal edge number of the shortest paths, denoted as $maxEN$. Obviously, $maxEN$ is $nSync$'s lower bound. 

Figure~\ref{nFrontier-benchmark} and ~\ref{nSync-benchmark} report the evaluation results on the 9 benchmark graphs. From two perspectives, the results showed effectiveness of the dynamic stepping. First, repeated relaxations are reduced significantly for graphs where redundant edge relaxations dominate runtime, and the degree distribution of graphs has almost negligible impact on the optimization effectiveness. On each graph except Road, the $averageEN$ and $maxEN$ show that the synchronization number can be reduced to the vertex number's logarithm, indicating that the runtime is dominated by redundant edge relaxations. On these graphs, the $nFrontier$ results are 1.01\;\~\;1.10; and the $nSync$ results are 1.55\;\~\;6.13. The results showed that most of the extended paths are the shortest paths and the synchronization number is only a few times of the vertex number's logarithm. Second, synchronization number is almost minimized for graphs where synchronizations dominate runtime. The $averageEN$ and $maxEN$ on Road showed that a shortest path averagely consists of more than 5 thousands of edges, indicating that the runtime is dominated by synchronizations. The results on Road showed that the achieved $nSync$ is very close to the graph's $maxEN$.

The $nFrontier$ results and the $nSync$ results on the 64 variant graphs are reported in Figure~\ref{nFrontier-variant} and ~\ref{nSync-variant} respectively. On 61 variant graphs, the $nFrontier$ results are less than 1.05, indicating that most of the extended paths are the shortest paths. The worst $nFrontier$ result is less than 1.20. On different graphs, the $nSync$ results are 0.54\;\~\;3.64. The results in Figure~\ref{nFrontier-variant} and ~\ref{nSync-variant} shows that the weight distribution of graphs has almost negligible impact on the dynamic stepping heuristic's optimization effectiveness.

Conclusively, the dynamic stepping enables the heuristic SSSP algorithm to significantly reduce repeated relaxations on graphs where redundant edge relaxations dominate runtime, while the synchronization cost is no more than a few times of the vertex number's logarithm. On most low-diameter graphs (69/72), more than 95 percents (1/1.05) of the extended paths are shortest paths, and the synchronization number is no more than a few times of the edge number's logarithm. The degree distribution and weight distribution of graphs have almost negligible impact on the dynamic stepping's optimization effectiveness. Furthermore, this heuristic is relatively insensitive to $\beta$ when it is within [0.85, 0.95].

\subsection{Evaluation Results of the Workload}
Combined with the dynamic stepping, the traversal optimization helps the EIC method to reduce an SSSP computation's workload effectively. The workload is measured with $nTrav$. As the dynamic stepping indicates that the $nTrav$ achieved on $\langle V, E, w\rangle$ is related with skewness of the graph's degree distribution, the $nTrav$ is compared with both $|E|/|V|$ and the degree distribution skewness. In the results, the skewness is measured by $log_{2}{(\max\limits_{x\in V}|adj(x)|/ highD(0))}$ and is denoted as $DD\_skewness$. Larger is $DD\_skewness$, more skewed is the degree distribution.

\begin{figure}[!t]
\begin{minipage}[c]{0.48\textwidth}
\centering
\subfloat[On the original graphs]{\includegraphics[width=0.45\textwidth]{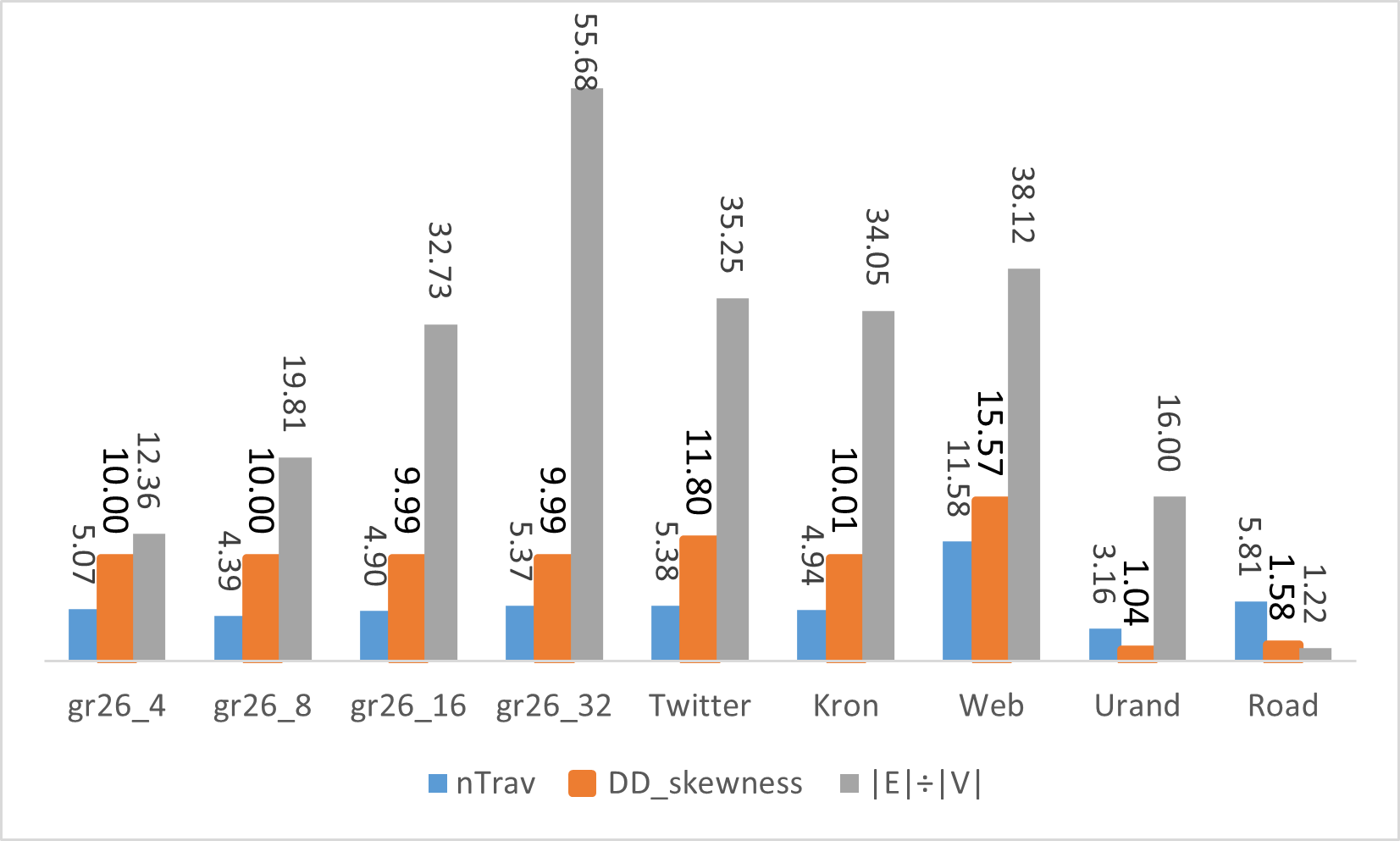}
\label{nVisit-original}}
\hfil
\subfloat[On the variant graphs]{\includegraphics[width=0.45\textwidth]{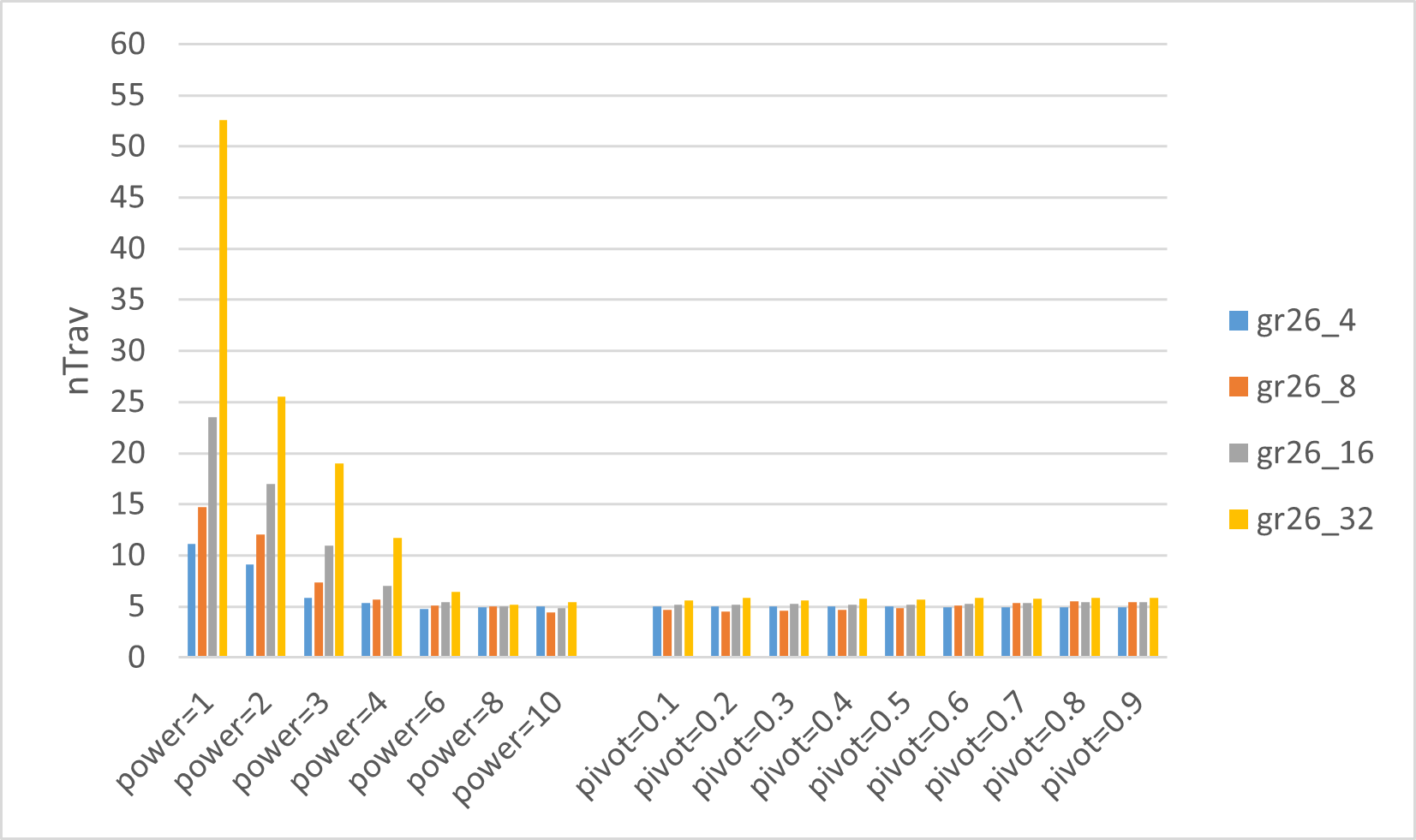}
\label{nVisit-variant}}
\end{minipage}
\begin{minipage}[t]{0.48\textwidth}
\centering
\caption{Evaluation results:  \emph{nTrav}.} 
\label{nVisit}
\end{minipage}
\end{figure}

The experimental results on the 9 benchmark graphs are reported in Figure~\ref{nVisit-original}. The results showed that the achieved $nTrav$ is less than $|E|/|V|/2$ if there are enough skippable edges. Such a graph's $nTrav$ tends to increase with the \emph{DD\_skewness}, and its $|E|/|V| - nTrav$ tends to increase with $|E|/|V|$. On each graph except Road, $|E|/|V| - nTrav$ is greater than 7.29. The best $|E|/|V| - nTrav$ is more than 50. The worst $nTrav$ is less than one third of the graph's $|E|/|V|$, and it was achieved on the graph with the largest $DD\_skewness$. On each graph except Web and Road, the $nTrav$ result is 3.16\;\~\;5.38. Web's $DD\_skewness$ is 15.57, its $nTrav$ result is 11.58. Road's $|E|/|V|$ is 1.22, its $nTrav$ result is 5.81. 

Figure~\ref{nVisit-variant} reports the results on the 64 variant graphs. The results showed that the $nTrav$ achieved on $\langle V, E, w\rangle$ tends to increase with $|E|/|V|$ and decrease with $|\{w(e): e\in E\}|$. The worst $nTrav$ was achieved on a variant graph created by modifying gr26\_32's weights with the integer ${power}=1$. In such a graph, different edges share the same weight, and there is more than one shortest path linking the same pair of vertices. On the variant graphs with at least 255 different edge weights, the $nTrav$ results are about 5.  

The results in Figure~\ref{nVisit} showed that EIC is more efficient than Dijkstra's algorithm when there are enough skippable edges. Dijkstra's algorithm traverses every edge reachable from the source vertex, and its workload on $\langle V, E, w\rangle$ can be denoted with $|E|/|V|$. On every graph except Road, EIC's $nTrav$ is less than the graph's $|E|/|V|/2$. EIC tends to be more efficient than Dijkstra's algorithm as the graph's $|\{w(e): e\in E\}|$ increases.

\subsection{Evaluation Results of the Flow-Controlling Overhead and Load-Balance}

\begin{figure}[!t]
\begin{minipage}[c]{0.48\textwidth}
\centering
\subfloat[On the benchmark graphs]{\includegraphics[width=0.45\textwidth]{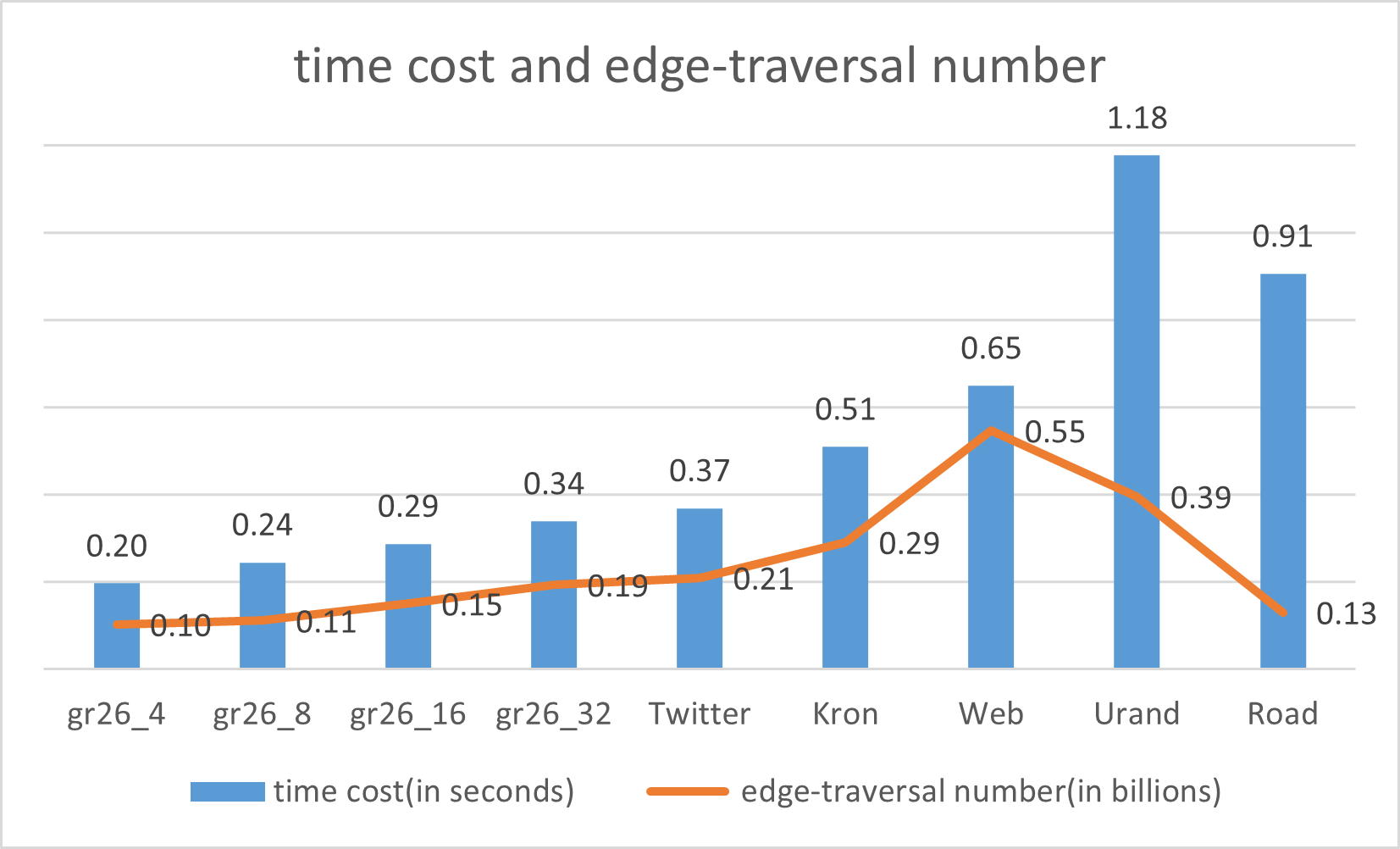}
\label{time-vs-nvisit-original}}
\end{minipage}
\begin{minipage}[c]{0.48\textwidth}
\centering
\subfloat[On the variant graphs]{\includegraphics[width=0.45\textwidth]{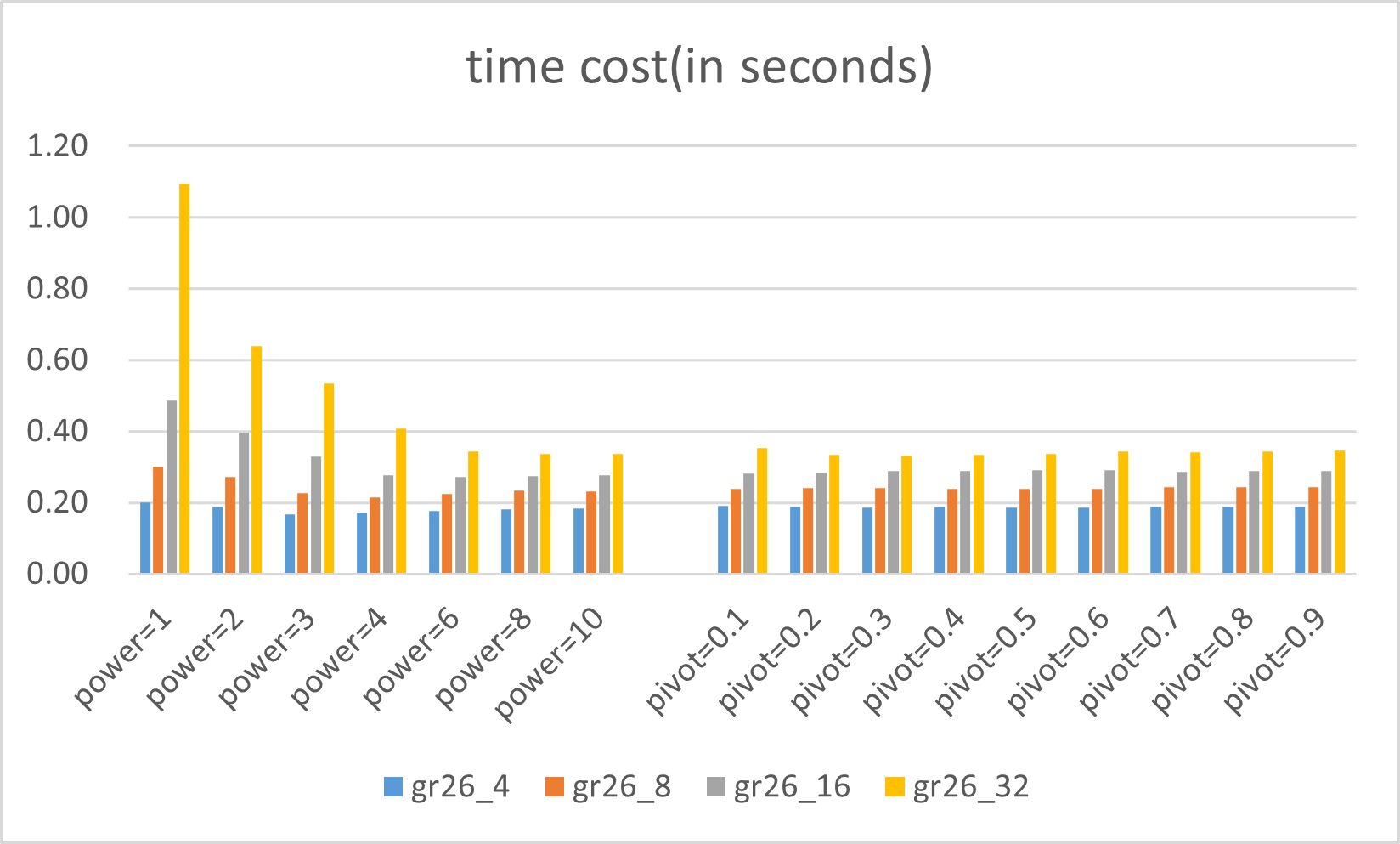}
\label{time-variant}}
\hfil
\subfloat[On the variant graphs]{\includegraphics[width=0.45\textwidth]{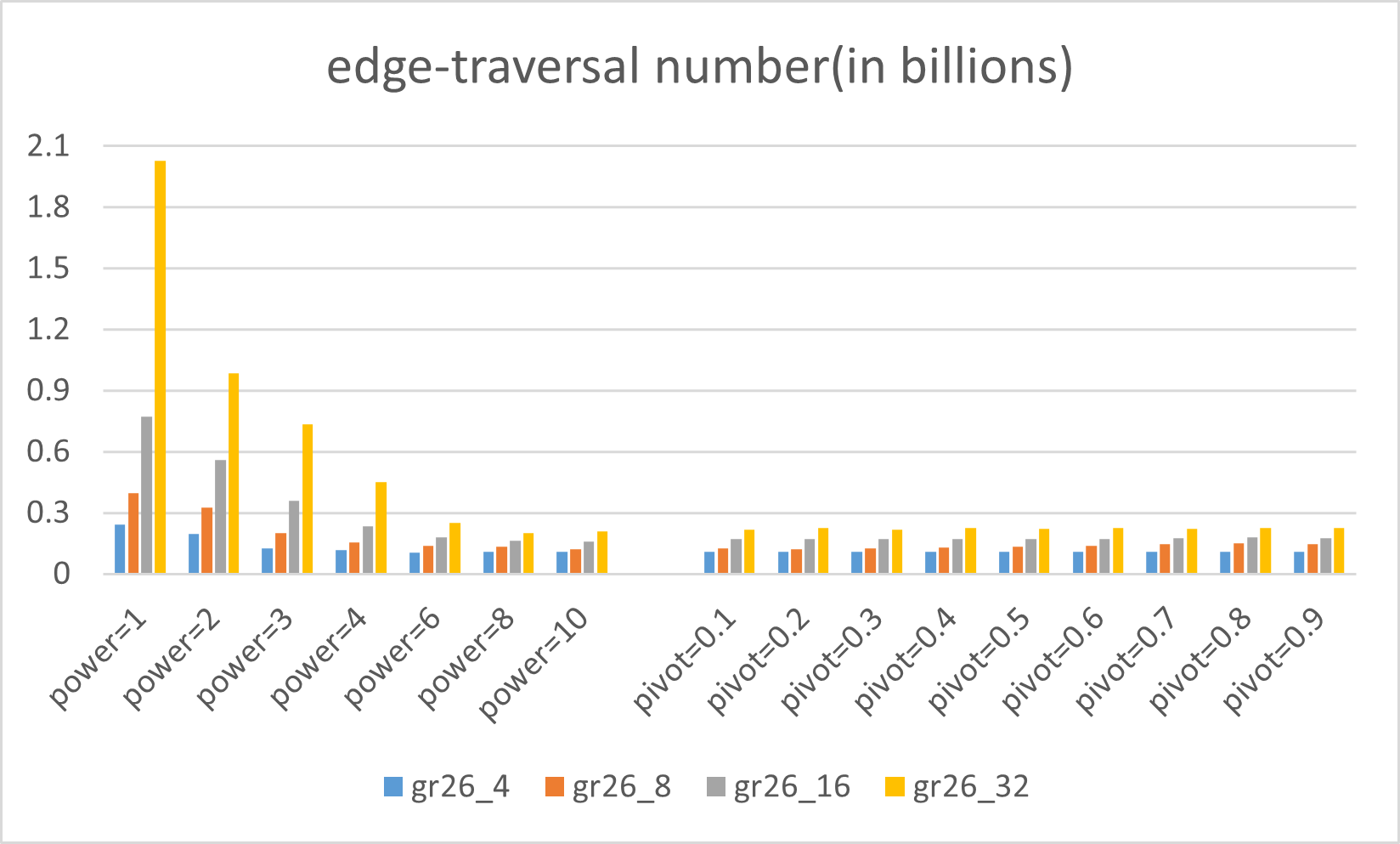}
\label{time-edge-visit}}
\end{minipage}
\begin{minipage}[t]{0.48\textwidth}
\centering
\caption{Evaluation results:  time cost and edge traversal number.} 
\label{time}
\end{minipage}
\end{figure}

The heuristic algorithm's overhead in parallel settings is evaluted by comparing its edge traversal number with the time cost. The overhead is mainly contributed by the synchronizations, the flow-controlling overhead, and the load imbalance. The results are reported in Figure~\ref{time}. On the graphs whose degree distributions are skewed, the time costs tend to be linearly proportional to the edge-traversal numbers. The results showed that EIC's overhead is almost ignorable if the graph's degree distribution is skewed enough.

Figure~\ref{time-vs-nvisit-original} reports results on the 9 benchmark graphs. Time costs and edge-traversal numbers on the 64 variant graphs are reported in Figure~\ref{time-variant} and Figure~\ref{time-edge-visit} respectively. Except for Urand and Road, each other graph's degree distribution is skewed. The results on these skewed graphs showed that EIC's overhead on such a graph is almost ignorable. 

The time cost on Urand is 1.18 seconds, while edge-traversal number is 0.39 billions. By comparing the results of Urand and Web, we believe that the extra time cost on Urand was contributed by the load imbalance. Urand's edge-traversal number is 0.39 billions, which is much less than Web's 0.55 billions. Results in Figure~\ref{nSync-benchmark} indicate that Web's overhead of synchronizations and flow-control are greater than that of Urand, since Web's synchronization number $6.13\times\log_2(50.6\times10^6)=157$ is greater than Urand's synchronization number $5.33\times\log_2(134.2\times10^6)=144$. In addition, the achieved results are also consistent with that the degrees of most vertices in Urand are $highD(0)$, as this characteristic indicates that only relatively a few of vertices were settled in each of the first steps.

Road's time cost is is 0.91 seconds. Compared with Web's results, more time were consumed on Road, while the edge-traversal number is less than one fourth of Web's. Combined with the results in Figure~\ref{nSync-benchmark}, we believe that the extra time cost was mainly contributed by the synchronizations. Road's synchronization number is $470.01\times\log_2(23.9\times10^6)=11546$. A synchronization is averagely shared by only about 10 thousand edge traversals. 

\subsection{Performance Comparison Results}
EIC was compared with five state-of-the-art SSSP implementations listed in Table~\ref{tab-implementations}. On each GAPBS benchmark graph, different implementations were compared by their time costs, \emph{nFrontier} results and \emph{nSync} results. PQ-$\Delta^*$'s results were achieved with its default $\Delta$($=2^{21}$). PQ-$\rho$'s results were achieved with its default $\rho$($=2^{21}$). Graph500's results were achieved with $\Delta=0.1\times maxW(G,1)$. GAPBS's $\Delta$ parameter was varied for each input graph, so as to achieve its best performance. The \emph{nFrontier} results and \emph{nSync} results are not reported for PQ-BF, PQ-$\Delta^*$ and PQ-$\rho$, because these three implementations have outputed only the time costs. 

It is notable that the bucket fusion optimization has been exploited in GAPBS, PQ-$\Delta^*$ and PQ-$\rho$ to reduce the synchronizations. In GAPBS, a thread's local bucket may have been updated more than one time before its vertices are merged into the global bucket, and the \emph{nFrontier} has counted only the vertices merged into the global bucket. Therefore, GAPBS's \emph{nFrontier} result is usually less than the number of the actually extended paths, and its \emph{nSync} result can be less than 1.

\begin{table}[!t]
\footnotesize
\caption{Five state-of-the-art SSSP implementations}
\label{tab-implementations}
\tabcolsep 4.0pt 
\begin{tabular*}{0.47\textwidth}{l|rr}
\toprule
                   &implemented algorithm&programming language\\\hline
Graph500~\cite{6}   & $\Delta$-stepping algorithm      & MPI, C/C++\\
GAPBS~\cite{7}      & $\Delta$-stepping algorithm      & OpenMP, C/C++\\
PQ-$\Delta^*$~\cite{19}& $\Delta^*$-stepping algorithm     &  CilkPlus, C/C++\\
PQ-$\rho$ ~\cite{19}  & $\rho$-stepping algorithm        &  CilkPlus, C/C++\\
PQ-BF~\cite{19}         & Bellman-Ford's algorithm&  CilkPlus, C/C++\\
\bottomrule
\end{tabular*}
\end{table}

Table~\ref{tab-gapbs} lists each implementation's time costs. On each GAPBS benchmark graph, EIC's time cost includes the preprocess time and the seach time. The preprocess time was consumed for computing the graph's \emph{RtoW[]} and sorting each vertex's incident edges. It was shared by different SSSP computations on the graph. The search time was consumed for runing the heuristic SSSP algorithm. 

\begin{figure}[!t]
\begin{minipage}[c]{0.48\textwidth}
\centering
\subfloat[{}]{\includegraphics[width=0.55\textwidth]{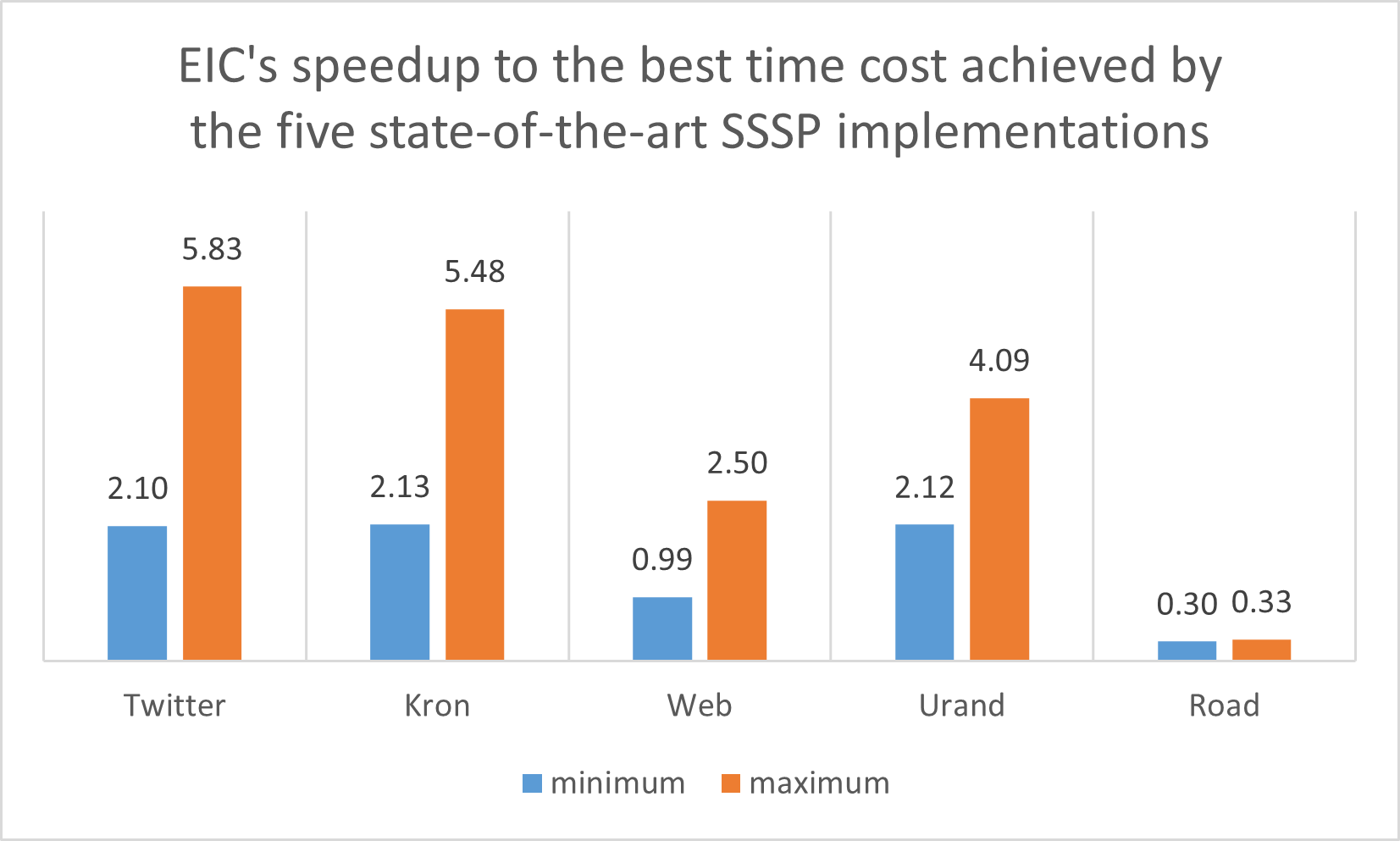}
\label{speedup-gapbs}}
\hfil
\subfloat[{}]{\includegraphics[width=0.45\textwidth]{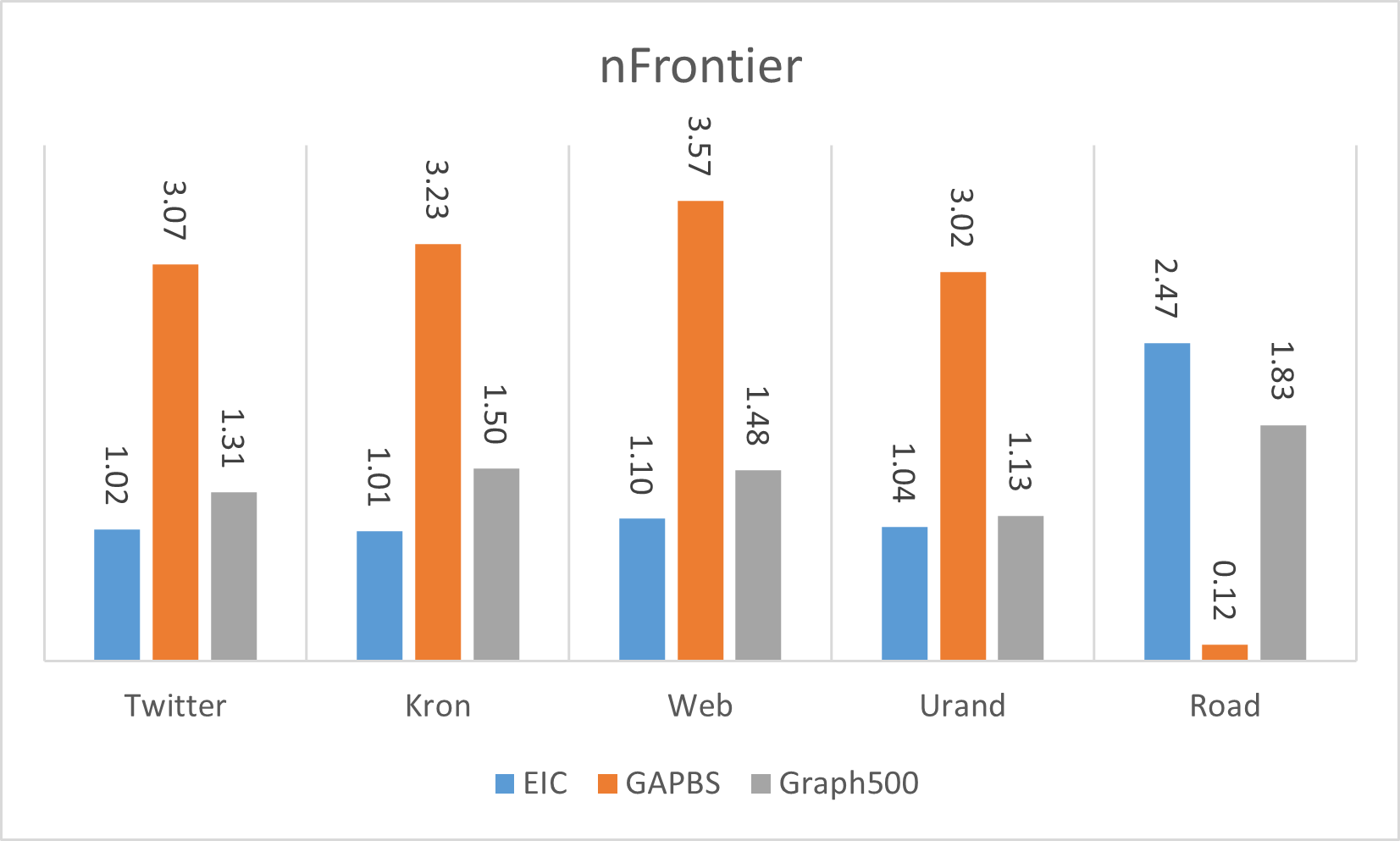}
\label{nFrontier-gapbs}}
\hfil
\subfloat[{}]{\includegraphics[width=0.45\textwidth]{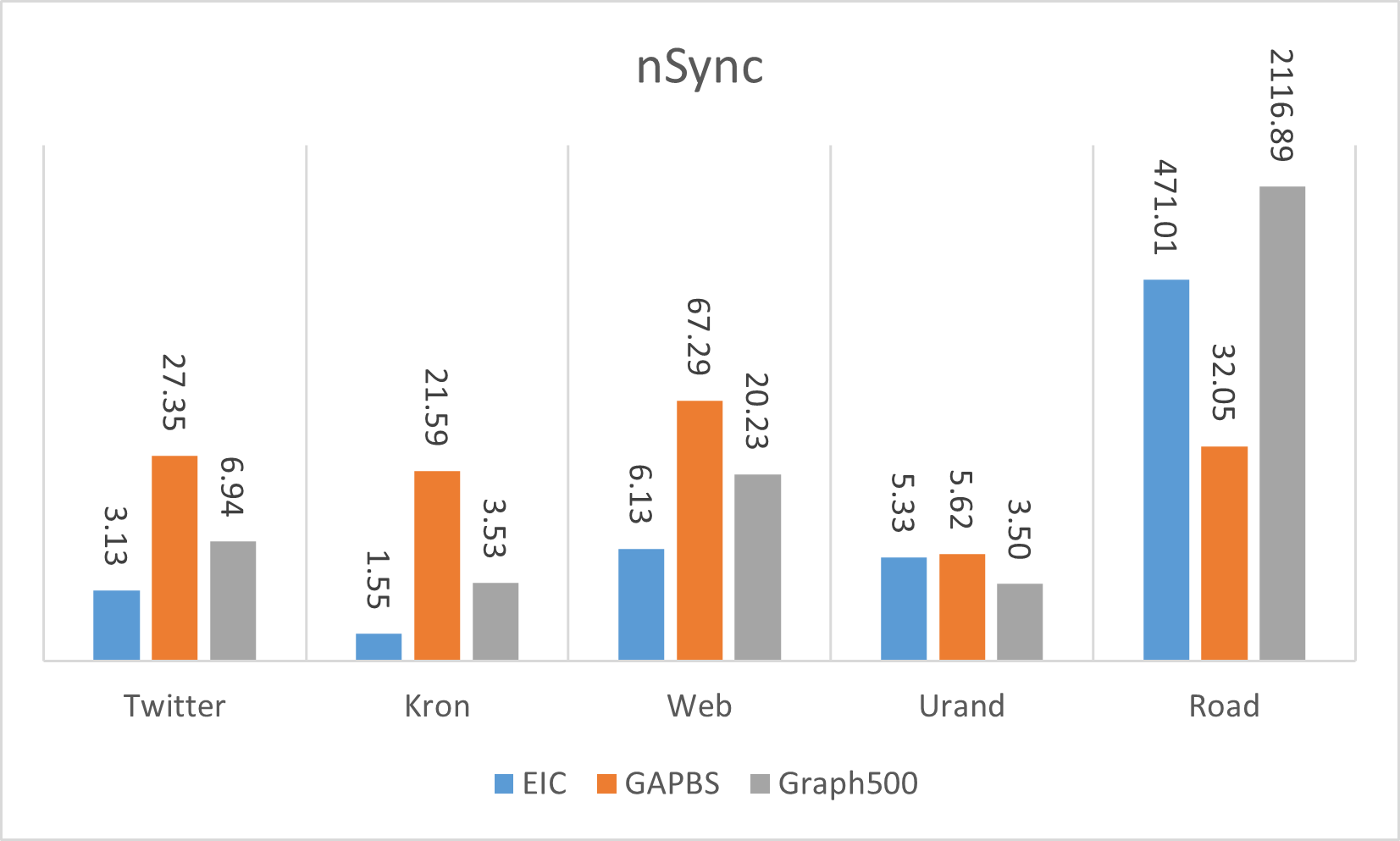}
\label{nSync-gapbs}}
\end{minipage}
\begin{minipage}[t]{0.48\textwidth}
\centering
\caption{Comparison on GAPBS benchmark graphs.} 
\label{Comparison}
\end{minipage}
\end{figure}

The comparison results are reported in Figure~\ref{Comparison}. We computed a minmum speedup and a maximum speedup for EIC. Each speedup is EIC's time cost to the best time cost achieved by the five state-of-the-art SSSP implementations. If there is only one SSSP computation on the graph, EIC achieves its minimal speedup, since its time cost is the sum of the preprocess time and the search time. When there are enough SSSP computations on the same graph, EIC achieves its maximal speedup, since the preprocess time's contribution to EIC's time cost is almost ignorable.

\begin{table}[!t]
\footnotesize
\caption{Time costs (in seconds)}
\label{tab-gapbs}
\tabcolsep 4.0pt 
\begin{tabular*}{0.47\textwidth}{l|rrrrr}
\toprule
        &Twitter&Kron&Web&Urand&Road\\\hline
GAPBS&2.147&2.790&1.954&4.816&0.300\\
		& $(\Delta=1)$ & $(\Delta=1)$ & $(\Delta=1)$& $(\Delta=3)$&$(\Delta=5400)$\\\hline
PQ\_BF&2.405&3.257&1.621&14.058&1.003\\
PQ\_$\rho$&2.198&3.480&1.843&5.146&1.025\\
PQ\_$\Delta^*$&2.266&3.488&1.628&13.496&1.029\\\hline
EIC\_search &0.368&0.509&0.649&1.177&0.906\\
EIC\_preprocess&0.654&0.803&0.985&1.097&0.104\\\hline
Graph500&3.803&5.610&4.966&7.095&7.160\\
\bottomrule
\end{tabular*}
\end{table}

On each graph except for Road, EIC achieved a much better performance than the five state-of-the-art SSSP implementations. Comparing with the best time cost achieved by the five state-of-the-art SSSP implementations, its maximal speedup is $2.50\times$ to $5.83\times$, and its minimal speedup is $0.99\times$ to $2.13\times$, as illustrated in Figure~\ref{speedup-gapbs}. From the results reported in Figure~\ref{nFrontier-gapbs} and ~\ref{nSync-gapbs}, we can conclude that the performance improvement is mainly contributed by the reduction in redundant relaxations and synchronizations. EIC's \emph{nFrontier} is only about one third of that achieved by GAPBs, and its synchronizations are also much less than that of Graph500 and GAPBS if the graph's degree distribution is skewed enough.


On Road, GAPBS achieved the best time cost, the best \emph{nFrontier} and the best \emph{nSync}. As shown in Figure~\ref{nSync-gapbs}, GAPBS's synchronizations are much less than the maximum of edges in shortest paths, resulting that its synchronization overhead is much less than that of other implementations. GAPBS's \emph{nFrontier} on Road also shows that its reported \emph{nFrontier} is much less than the number of the actually extended paths.

\section{Conclusion}
This paper presents our work for heuristically optimizing SSSP computations on undirected graphs. We have introduced the EIC method. It is a novel shortest path search method that uses length threshod pairs to index and classify each vertex's adjacent edges. By partitioning each vertex's adjacent edges into irrelevant edges, long relevant edges and short relevant edges, this method can significantly reduce redundant edge relaxations at the cost of almost negligible synchronization overhead. We have proposed the dynamic stepping heuristic and the the traversal optimization heuristic. The dynamic stepping helps the EIC method to reduce repeated relaxations. The traversal optimization helps the EIC method to reduce edge traversals for shortest paths that are created by relaxations of the long relevant edges. Both heuristics work with the graph's vertex-degree statistics and edge-weight statistics.

Based on the two heuristics, the EIC method is used to develop a heuristic SSSP algorithm. It uses the dynamic stepping to construct a set of scheduling thresholds pairs. The traversal optimization is then used to choose a selection threshold for each pair. With these pairs and the selection thresholds, each vertex's adjacent edges are heuristically indexed and classified, and the the SSSP computation is decomposed into concurrent edge traversals. These pairs are also used to schedule the edge traversals. Due to the dynamic stepping, this heuristic SSSP algorithm is almost comparable to Dijkstra's algorithm in reducing repeated relaxations for low-diameter graphs, while the extra synchronization overhead is almost negligible. Furthermore, this algorithm employs the traversal optimization to skip more edges with less cost than that in~\cite{34}.

The two heuristics and the algorithm were evaluated on 73 real-world and synthetic graphs. These graphs span diverse vertex degree distributions and edge weight distributions. On most low-diameter graphs, more than 95 percents of the extended paths are shortest paths, the synchronization number is no more than a few times of the edge number's logarithm, and the edge traversal number is less than half the edge number. The algorithm was also compared with five state-of-the-art SSSP implementations. On each GAPBS benchmark graph except Road, the algorithm's speedup to the best time achieved by these five state-of-the-art SSSP implementations is $2.50\times$ to $5.83\times$.

It is notable that the EIC method, dynamic stepping and traversal optimization are independent of the heuristic SSS algorithm. They can be used individually or in combination to solve other shortest path search problems like All-Pair-Shortest-Path and Point-to-Point Shortest Path. The heuristics also can be used individually or in combination to improve other algorithms like $\Delta$-stepping and their implementations.

\end{document}